\documentclass{jfm}
\usepackage{graphicx,color}
\usepackage{float}
\usepackage{epstopdf, epsfig}
\usepackage{caption,subcaption}
\usepackage[intlimits]{amsmath}
\usepackage{lscape}
\usepackage{rotating}
\usepackage{amssymb, amsmath}
\usepackage{amsmath}
\usepackage{xcolor}
\usepackage{hyperref} 

\newcommand{\MDG}[1]{{\textcolor{red}{*** #1 ***}}}
\newcommand{\MDGrevise}[1]{{\textcolor{black}{#1}}}

\newcommand{\CRCA}[1]{{\textcolor{black}{#1}}}

\usepackage[normalem]{ulem}

\newcommand{\Koop}{\mathcal{K}}
\newcommand{\Kapp}{\mathsfbi{K}}
\newcommand{\bh}{\boldsymbol{h}}
\newcommand{\bg}{\boldsymbol{g}}
\newcommand{\bx}{\boldsymbol{x}}

\shorttitle{Koopman for Shear Flows}
\shortauthor{C. R. Constante-Amores and others}

\title{
Data-driven Koopman operator predictions of turbulent dynamics in models of shear flows
}

\author{C. Ricardo Constante-Amores$^{1,2}$,
Andrew J. Fox$^{1}$, Carlos E. P\'{e}rez De Jes\'{u}s$^{1}$, and Michael D. Graham$^{1}$ \corresp{\email{mdgraham@wisc.edu}}%
}
  
\affiliation{
$^1$Department of Chemical and Biological Engineering, University of Wisconsin-Madison, Madison WI
53706, USA\\
$^2$Department of Mechanical Science and Engineering, University of Illinois, Urbana Champaign, IL 61801, USA
}
\begin{document}

\maketitle

\renewcommand{\sout}[1]{\unskip}

\begin{abstract}
The Koopman operator enables the analysis of nonlinear dynamical systems through a linear perspective by describing time evolution in the infinite-dimensional space of observables. Here this formalism is applied to shear flows, specifically the minimal flow unit plane Couette flow and a nine-dimensional model of turbulent shear flow between infinite parallel free-slip walls under a sinusoidal body force (Moehlis et al., New J. Phys. 6, 56 (2004)). We identify a finite set of observables and approximate the Koopman operator using neural networks, following the method developed by Constante-Amores et al., Chaos, 34(4), 043119 (2024). Then the time evolution is determined with a method here denoted as “Projected Koopman Dynamics”. Using a high-dimensional approximate Koopman operator directly for the plane Couette system is computationally infeasible due to the high state space dimension of direct numerical simulations (DNS). However, the long-term dynamics of the dissipative Navier-Stokes equations are expected to live on a lower-dimensional invariant manifold. Motivated by this fact, we perform dimension reduction on the DNS data with a variant of the IRMAE-WD (implicit rank-minimizing autoencoder-weight decay) autoencoder  (Zeng et al., ML\&ST 5(2), 025053 (2024)) before applying the Koopman approach. We compare the performance of data-driven state space (neural ordinary differential equation) and Projected Koopman Dynamics approaches in manifold coordinates for plane Couette flow. Projected Koopman Dynamics effectively captures the chaotic behavior of Couette flow and the nine-dimensional model, excelling in both short-term tracking and long-term statistical analysis, outperforming other frameworks applied to these shear flows.

\end{abstract}

\keywords{shear flows, Koopman, data-driven models, chaos}

\maketitle

\section{Introduction}

The study of turbulence, particularly in wall-bounded flows, holds paramount importance in understanding  fluid dynamics as it has profound implications for various engineering applications  \citep{Pope_2000,Avila}. 
\CRCA{Approximately $25\%$ of the energy consumed by industry  is dedicated to transporting fluids through pipes and channels, and about one-quarter of this energy is dissipated due to turbulence occurring near walls
}
\citep{jimenez_1991,waleffe_2001,Jimenez_2013}.
To understand the underlying physical mechanisms of turbulence, researchers have  focused on canonical flows such as plane Couette flow, which serves as an ideal test case  due to its simplicity and well-defined boundary conditions,
as evidenced by numerous studies,  such as \citet{nagata_1990,jimenez_1991,Waleffe_prl,KAWAHARA_KIDA_2001,waleffe_2001,Gibson,Itano}.

\MDGrevise{The focus of the present work is the extent to which modern data-driven methods can capture the dynamics of turbulent shear flows near transition.}
\CRCA{\sout{In this work, we study the dynamics of shear flow turbulence.}  Because of its geometric and dynamical simplicity, the  `minimal flow unit'  (MFU) of plane Couette flow (PCF) has been the object of much study \citet{jimenez_1991,Hamilton1995,Gibson,Gibson_2008,Gibson_2009,Halcrow_2009}.
The MFU is the minimum domain size in which turbulence will self-sustain; thus contains the key components of turbulent dynamics, i.e.~the `self-sustaining process' (SSP) \citep{Hamilton1995} in which low-speed streaks near the wall become wavy, triggering their breakdown to generate rolls, which in turn rolls lift fluid from the walls, regenerating the streaks. Given that the state space dimension (number of mesh points times number of velocity components) for fully resolved direct numerical simulation (DNS) even of an MFU may be $\mathcal{O}(10^5)$ \citep{jimenez_1991,Hamilton1995}}, significant efforts have been made to develop low-dimensional models capable of capturing the SSP 
\citep{Waleffe_1997,moehlis2004low,Gibson,Kaszas}. 
\citet{Waleffe_1997} proposed a 8-mode  model for turbulent shear flow between infinite parallel free-slip walls subjected to a sinusoidal body force. The modes were selected based on intuition from the self-sustaining process (SSP), and a Galerkin projection was then performed onto these modes.
\citet{moehlis2004low} proposed the
\CRCA{Moehlist, Faisst, and Eckhardt} 
(MFE) 
model, which contains an additional mode that enables modification of the mean profile by the turbulence; as the fluctuations change the mean profile for turbulent shear flows.
\sout{designed specifically for turbulent shear flow between infinite parallel free-slip walls subjected to a sinusoidal body force.}
The MFE model \sout{ has shown promise in faithfully representing the turbulent dynamics of the system} \MDGrevise{is attractive because it captures key features of shear flow transition near onset, including interaction between roll and streak modes, intermittency and ultimate relaminarization} while maintaining computational tractability. \MDGrevise{The present work addresses data-driven modeling of the plane Couette MFU both in the form of full DNS results and the MFE model.}
\CRCA{}


In data-driven modelling of  the dynamics, two different approaches can be taken: the state space approach and the function space or observable space approach. For the latter, the notion that any finite-dimensional non-linear dynamical system can be transformed into an equivalent linear system by lifting it to an infinite-dimensional space of  observables was introduced almost a century ago by \citet{koopman1931}. However, its significance emerged over a decade ago with its connection to dynamic mode decomposition (DMD), as outlined by \citet{schmid_2010}\CRCA{, and connected to the Koopman operator by \citet{Rowley}.}
\CRCA{ The non-linear dynamical system is transformed into a linear one using a set of observables, with the infinite-dimensional Koopman operator applying to all observables. Then,}
\sout{In DMD,} the entire dynamics is \sout{can be} \CRCA{represented}  by a linear expansion of the \sout{DMD} modes (eigenfunctions) and their corresponding  eigenvalues \citep{schmid2022dynamic,Mezic_arfm}. 
The challenge lies in lifting the state space to the observable space. 
\CRCA{In practice, the infinite-dimensional Koopman operator is approximated by a matrix-valued operator; this approximation aims to maintain the essential properties of the operator while operating only on a discrete finite set of observables.}
\CRCA{Several strategies have been proposed to ensure that the input function space, and subsequently the selection of  observables, is sufficiently `rich'. For a given method, there is one dictionary of observables, with many elements.}
\citet{eDMD} \CRCA{proposed the extended dynamic mode decomposition (EDMD), which uses }  Legendre polynomials of the state space coordinates \CRCA{as a  dictionary 
 to create the observable space}. Their approach primarily focused on low-dimensional systems, highlighting the difficulty of choosing appropriate dictionaries \citep{korda2018convergence}. \MDGrevise{Even for a simple dictionary such as orthogonal polynomials, the number of dictionary elements grows rapidly with the number of dimensions.}

\CRCA{Recently, attempts have been made to use neural networks to construct a good dictionary of  observables, resulting in a good approximation of the matrix-valued Koopman operator -- see \citet{Qianxiao,deepDMD,Lusch,lrans,deep_dmd,edmd_dl_ad}.} 
\CRCA{\citet{Qianxiao} presented extended dynamic mode decomposition with dictionary learning (EDMD-DL) which uses neural networks to adapt the dictionary of observables to the data.
}
\sout{proposed the idea of allowing neural networks to adapt to the data, resulting in a more robust method tailored to the system.} Recently, \citet{edmd_dl_ad} proposed a variant of the EDMD-DL that leverages automatic differentiation to simultaneously determine the dictionary of observables and the corresponding approximation of the Koopman operator. \MDGrevise{In their study, the authors contrast two different ways to handle temporal evolution, referred to as KDL$_{\mathrm{oo}}$ (Koopman Dictionary Learning, Observable-Only evolution) and KDL$_{\mathrm{so}}$ (State-Observable alternation). The former approach maintains the linearity of the Koopman formalism but suffers from poor performance in time evolution. The latter approach behaves substantially better in time evolution but is nonlinear. We use the latter approach here, denoting it as `Projected Koopman Dynamics'.} 
\CRCA{Finally, \citet{pan2024lifting} emphasized the importance of considering symmetries when learning the Koopman operator for low-dimensional systems.}

\CRCA{The accurate simulation of MFU PCF requires a large state space to resolve all the relevant spatial and temporal scales. For example, \citet{alec_couette,Hamilton1995} required approximately $\mathcal{O}(10^5)$ degrees of freedom to capture the complex nonlinear turbulent dynamics. Performing Koopman modeling in this state space poses computational challenges, as it involves lifting this high-dimensional state space into even a higher-dimensional space of observables.  }
\sout{Dealing with high-dimensional systems poses computational challenges when attempting to lift a finite-dimensional nonlinear state space into an infinite-dimensional linear space using any type of dictionary elements.}
\CRCA{However, the Navier-Stokes equations  (NSE) are dissipative PDEs; as a result, the long-term dynamics of the system are confined within a finite-dimensional invariant manifold with fewer degrees of freedom than the full state-space dimension\citep{Temam,Zelik}.}
\sout{addressed this issue  in the study of high-dimensional state space for pipe flow  and Kolmogorov flow. They modeled turbulence from a dynamical systems perspective, leveraging the dissipative nature of the Navier-Stokes equations  (NSE).  As a result, the long-term dynamics of the system are confined within a finite-dimensional invariant manifold with fewer degrees of freedom.} 
\MDGrevise{Thus if a coordinate representation (i.e.~a dimension-reduction process) can be found for data on an invariant manifold,  
then one can } lift the coordinates of this invariant manifold into its equivalent \sout{infinite-dimensional space} \CRCA{observable-space representation}, \CRCA{as was shown by \citet{pipe_flow} for the Koopman modelling of  coherent state dynamics on invariant manifolds.} \CRCA{Since in the present work, we are considering time evolution with Koopman operators on invariant manifolds, it is worth noting that \citet{Nakao} and \citet{mezic2020koopman} showed that these correspond to joint zero level sets of Koopman eigenfunctions.}

\MDGrevise{The dimension of an invariant manifold for long-time dynamics is not known a priori, so must be estimated from data.}
Various methods for estimating manifold dimensions have been introduced in the literature. For the Kuramoto-Sivashinsky equation (KSE), the manifold dimension has been estimated using covariant Lyapunov vectors and  monitoring the drop of the Lyapunov spectrum \citep{Yang,Yang_2009}. These methods require access to the governing equations and high-precision solutions, which are generally not available for the NSE. When the governing equations of motion are unknown and only data is available, the most common method is Principal Component Analysis (PCA, also known as Proper Orthogonal Decomposition, POD, in the fluid dynamics community) which is based on the idea of the projection of the state onto the set of orthogonal modes that capture the maximum energy of the data
\citep{Jolliffe1986, abdi2010principal,Holmes_2012}.
PCA, being a linear reduction method, \CRCA{projects data onto a flat manifold, which is not generally appropriate for complex nonlinear problems, and subsequently, it} struggles to estimate their true dimension.
\CRCA{Thus, nonlinear methods can be advantageous. One of the prevalent methods for nonlinear dimension reduction is the use of an undercomple autoencoder,  a pair of neural networks in which one network maps from a high-dimensional space to a low-dimensional space, and the other maps back \citep{Kramer,Hinton,Milano}. }
\MDGrevise{For very high-dimensional systems, it can be advantageous to perform an initial linear dimension reduction step with PCA, then use an autoencoder for further dimension reduction \citep{alec_pre,Young_simple}
} 
\MDGrevise{It can also be advantageous to apply PCA and an autoencoder in parallel to learn a low-dimensional nonlinear representation as a correction to a linear representation  \citet{alec_pre}.}\sout{
The authors studied the 1D KSE with a domain size of $L=22$ and observed a several orders-of-magnitude drop in the mean-squared error (MSE) of the data reconstruction once the dimension of the inertial manifold was reached.} Recently, \citet{jing2020implicit} introduced a variation of a standard autoencoder 
to learn low-rank representations for image-based classification. This method was extended to \MDGrevise{data from} dynamical systems by \citet{irmae}, demonstrating that this architecture can yield robust and precise estimates of manifold dimension, as well as an orthogonal manifold coordinate system. \MDGrevise{The present work uses a variation of this approach.}

Once, we have access to the manifold coordinates, i.e.~ a low-dimensional representation of the data, we can \sout{combine it 
with classical techniques
 \MDG{but we are not doing that.} for time evolution.} 
\CRCA{learn an evolution equation in these coordinates. Some examples where a state space modelling in manifold coordinates for chaotic systems include the KSE equation \citep{lrans,alec_pre,ssm_chaos}, Kolmogorov flow \citet{carlos_prf} and Couette flow  \citet{Kaszas,alec_chaos}. }
\citet{alec_pre} introduced an invariant-manifold-based data-driven approach known as `DManD' (data-driven manifold dynamics). DManD is based on the idea of performing a nonlinear reduction of the full space with autoencoders to find a mapping to the manifold coordinates, and then use a  neural ODEs (NODE)  to learn an evolution equation in manifold coordinates -- in NODE, a neural network learned from data parameterises the vector field (right-hand side of the ODE) determining the dynamics \citep{alec_chaos,chen2019neural}. This framework  applied to MFU PCF only requires  18 degrees of freedom to build reduced-order models that faithfully capture the turbulent dynamics over both short and long timescales, including the streak breakdown and regeneration process and Reynolds stresses \citep{alec_couette}. \CRCA{We remark that NODE is an advantageous approach as the dynamics are Markovian (consistent with dynamics that arise from ODEs) and continuous in time.}

\CRCA{We conclude our introduction by discussing several works in data-driven modeling of the MFE system.  \citet{Racca2022}  modeled MFE using Echo State Networks, which expand the nonlinear input state space to create a high-dimensional system (up to two orders of magnitude higher). Their predictions account for history in their time evolution, making the model non-Markovian, with temporal evolution based on discrete time maps. \citet{Eivazi} employed a Koopman-based framework, proposed by \citet{khodkar2019koopman}, incorporating a nonlinear forcing term. They construct the forcing term as a library of candidate nonlinear functions, allowing the model to identify the most significant nonlinearities from the data. Here, the inclusion of  predetermined forcing candidates is needed beforehand. Their method can capture accurate short-term predictions and reproduce long-term statistics, but the temporal evolution includes delay embedding, thus rendering the model non-Markovian. \MDGrevise{Similarly, approaches based on transformer architectures have been successful for prediction of flow problems, but are also non-Markovian \citep{Solera-Rico.2024.10.1038/s41467-024-45578-4}.} Finally, \citet{Fox2023} utilise a state space approach  denoted as Charts and Atlases for Nonlinear Data-Driven Dynamics on Manifolds, or CANDyMan \citep{Floryan}. In their approach, the time series of MFE data is divided into overlapping clusters or charts, and an evolution equation is learned for each chart to construct a final global model. This clustering of the data significantly improved the performance of the models.}

\begin{figure}
\centering
\begin{tabular}{cc}
\includegraphics[width=0.5\textwidth]{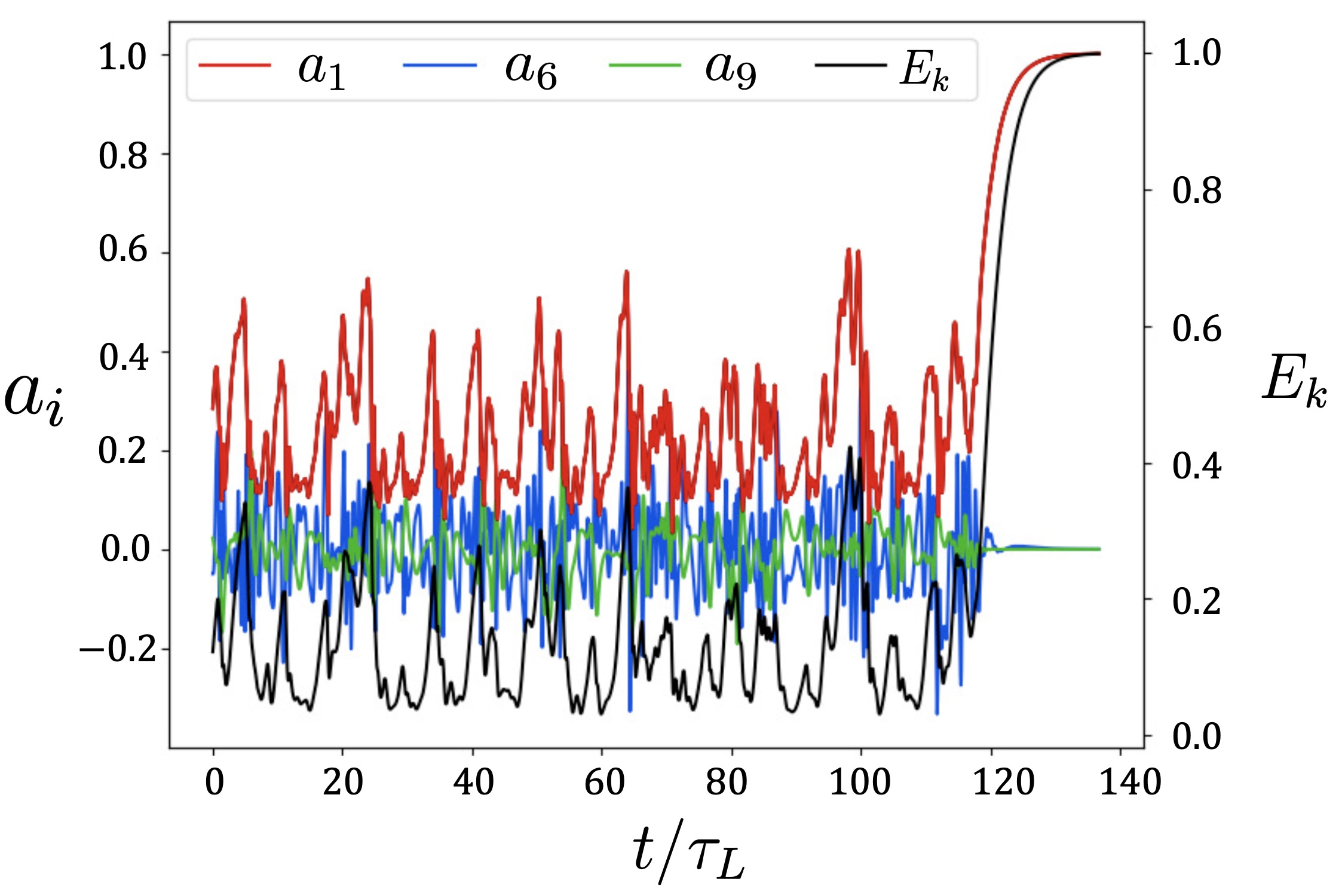}&
\includegraphics[width=0.4\textwidth]{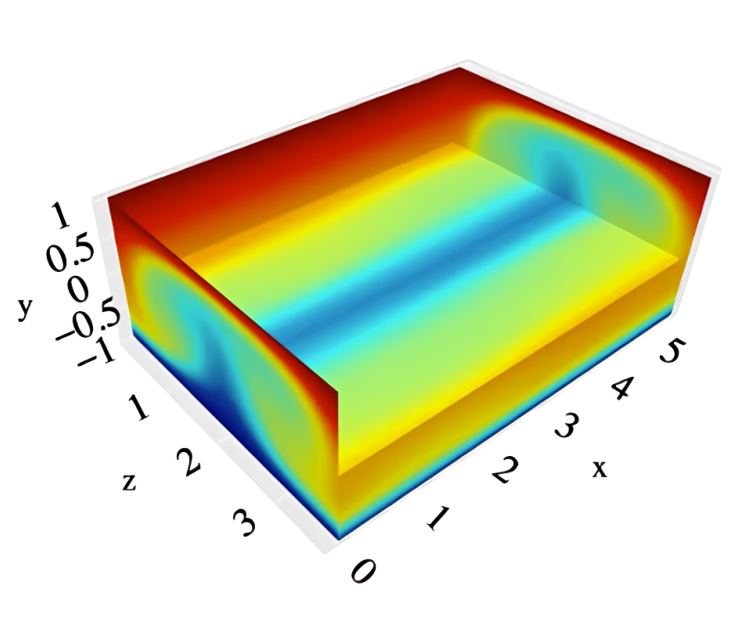}\\
(a) & (b)\\
\end{tabular}
\caption{Schematic representation of the shear flows considered in this study.
(a) Evolution of three amplitudes, $a_1, a_6, a_9$ and corresponding kinetic energy from one time series of the MFE data set.
(b) Three dimensional snapshot of turbulent MFU Plane Couette Flow at $Re=400$, coloured by the streamwise velocity.
}
\label{intro}
\end{figure}

\CRCA{In this study, we aim to build data-driven Koopman-operator-based predictions of turbulent dynamics in models of shear flows using the approach presented in \citet{edmd_dl_ad}. To achieve this, we begin with the nine-dimensional MFE model, and then we extend our framework to the high-dimensional  MFU PCF (see Figure \ref{intro}, for a schematic representation of the MFE model and MFU PCF) . We remark that the MFE model, being inherently low-dimensional, does not require dimension reduction.
In contrast, the plane Couette flow requires dimension reduction to manifold coordinates prior to lifting the system to the observable space.}
To find these manifold coordinates, we introduce a variant of IRMAE-WD. Then, we  employ a Koopman approach to learn an effective observable-space  
\sout{infinite-dimensional}
 representation and dynamical system using the framework that we termed as `Koopman-Data-Driven Manifold Dynamics' (KDManD). We also make comparisons with DManD (state space approach using NODEs).  The paper is organized as follows, Section \ref{framework} provides a detailed explanation of the framework \MDGrevise{for finding good observables for an approximate Koopman representation} with neural networks as well as the time-evolution of the observables, along with a description of the approach to find the manifold coordinates for PCF. Section \ref{Results} presents the description of the MFE model and PCF system, the results  concerning the reconstruction of short-time and long-time statistics. Finally, concluding remarks are summarized in Section \ref{sec:conclusions}.

\section{Framework \label{framework}}

\begin{figure}
\centering
\begin{tabular}{cc}
\includegraphics[width=\textwidth]{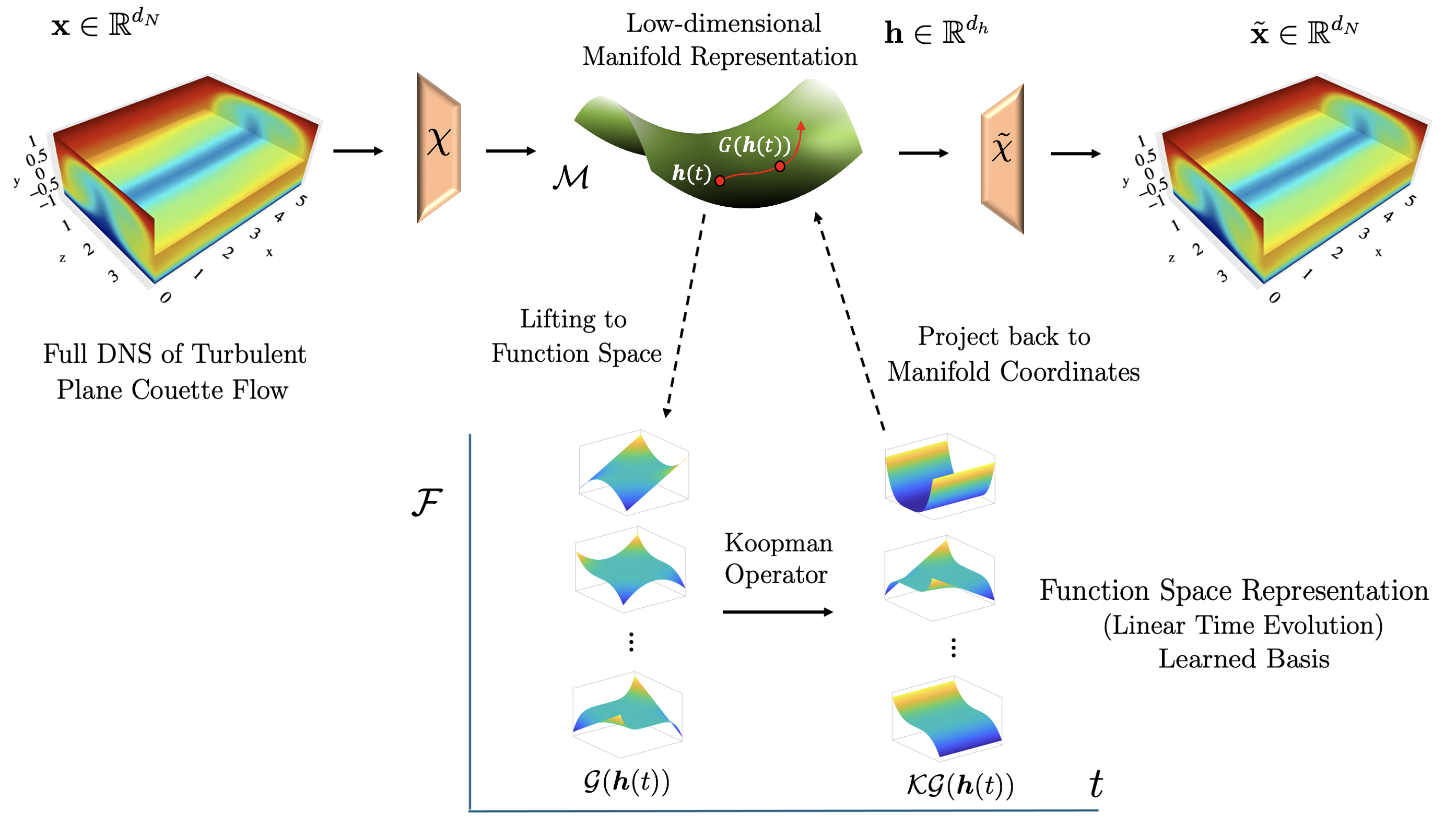}
\end{tabular}
\caption{A cartoon of the data-driven framework applied to the full DNS of turbulent plane Couette flow.
The top panel represents the methodology to find a coordinate transformation from the full state space ($\mathbf{x} \in \mathbb{R}^{d_N}$) to a low-dimensional representation ($\boldsymbol{h} \in \mathbb{R}^{d_h}$)  \CRCA{via $\boldsymbol{\chi}$ and $\tilde{\boldsymbol{\chi}}$}
\sout{an encoder ($\mathcal{E}$) and decoder ($\mathcal{D}$)}
In manifold coordinates, we can use a state space modeling approach to update the state $\boldsymbol{h} \in \mathcal{M}$ using the state space time-evolution operator $\boldsymbol{G}_{\delta t}$, i.e., 
\CRCA{$\boldsymbol{h}(t+\delta t)=\boldsymbol{G}_{\delta t}(\boldsymbol{h}(t))$}.
Alternatively, the manifold representation can be lifted to the high-dimensional space of observables where the dynamics behave linearly. The bottom path represents the observable space in which the updates the observables, $\mathcal{G}(\boldsymbol{h}) \in \mathcal{F}$, is done using the Koopman operator, $\mathcal{K}$, i.e., 
$\mathcal{G}(\boldsymbol{h}(t+\delta t))=\Koop\mathcal{G}(\boldsymbol{h}(t))$. As a result, either the state space or function space modeling \CRCA{leads to the same result}\sout{are commutative,}.
\sout{allowing one to choose between solving finite-dimensional but nonlinear problem (the top path) or an infinite-dimensional but linear \MDG{is the problem still linear when we project after every time step? } problem (the bottom path) using the Koopman operator. }\label{intro_cartoon}}
\end{figure}

Figure \ref{intro_cartoon} shows a diagram of the framework applied to a full DNS plane Couette flow, where the evolution of states and observables are done in manifold coordinates, which encapsulate the long-time dynamics as the full DNS.

\subsection{Finding manifold coordinates \label{irma_br_section}}

In this work, we consider systems that are characterized by deterministic, Markovian dynamics, so if $\boldsymbol{x} \in \mathbb{R}^{d_N}$
represents the full space state, then the dynamics can be represented by an ODE as
\begin{equation} \label{eq:ODE}
	\dfrac{d\boldsymbol{x}}{dt}=\boldsymbol{f}(\boldsymbol{x}).
\end{equation} 
Here, $\boldsymbol{x}$ represents the full state space 
from highly resolved direct numerical simulation (DNS). 
\CRCA{The long-time dynamics of dissipative systems collapse onto an invariant manifold of embedding dimension $d_h$ (i.e., $d_h \ll d_N)$. }
When $\boldsymbol{x}$ is mapped into the manifold coordinates, represented by $\boldsymbol{h}\in \mathbb{R}^{d_h}$, an equivalent evolution equation can be expressed as 
\begin{equation}
    \frac{d\boldsymbol{h}}{dt} = \boldsymbol{g}(\boldsymbol{h}).\label{eq:StateEvolution}
\end{equation}

With the full state, we use an undercomplete autoencoder architecture to find a mapping for the low-dimensional representation  as
$\boldsymbol{h}=\boldsymbol{\chi}(\boldsymbol{x})$, along with the inverse $\check{\boldsymbol{x}}=\check{\boldsymbol{\chi}}(\boldsymbol{h})$
so that the full state space may be recovered  (i.e., $\boldsymbol{x} \approx \tilde{\boldsymbol{x}}$).

To find $\boldsymbol{\chi}$ and $\check{\boldsymbol{\chi}}$,  we use a variation of IRMAE-WD \citep{irmae}. 
\CRCA{In the original IRMAE-WD, the autoencoder is formed by standard nonlinear encoder, $\mathcal{E}$, and decoder, $\mathcal{D}$, networks with $n$ additional linear layers with weight matrices $\mathcal{W}_1,\mathcal{W}_2\cdots\mathcal{W}_n$ (of size $d_z\times d_z$).  The encoder finds a compact representation $\boldsymbol{z} \in \mathbb{R}^{d_z}$, and the decoder performs the inverse operation. In IRMAE-WD, a predetermined number of linear layers is specified in advance. The variant that we introduce in this research allows the forward pass to propagate through different `branches' with varying numbers of linear layers (e.g., an upper and lower range of linear layers are specified). We refer to this architecture as IRMAE-WD-B. This variant  improves the robustness of dimension estimates for more complex systems as presented by \citet{Carlos_phd}. }

\sout{The primary difference between the variant used in this research and IRMAE-WD is that the original IRMAE-WD architecture requires a predetermined number of linear layers\MDG{too soon to talk about this -- you haven't introduced the IRMAE-WD architecture at all yet} to be specified in advance. In contrast, our variant allows the forward pass to propagate through different branches with varying numbers of linear layers. We refer to this architecture as IRMAE-WD-B. This variant  improves robustness of dimension estimates for more complex systems as presented by } 

\begin{figure}
	\centering
	\includegraphics[width=\linewidth]{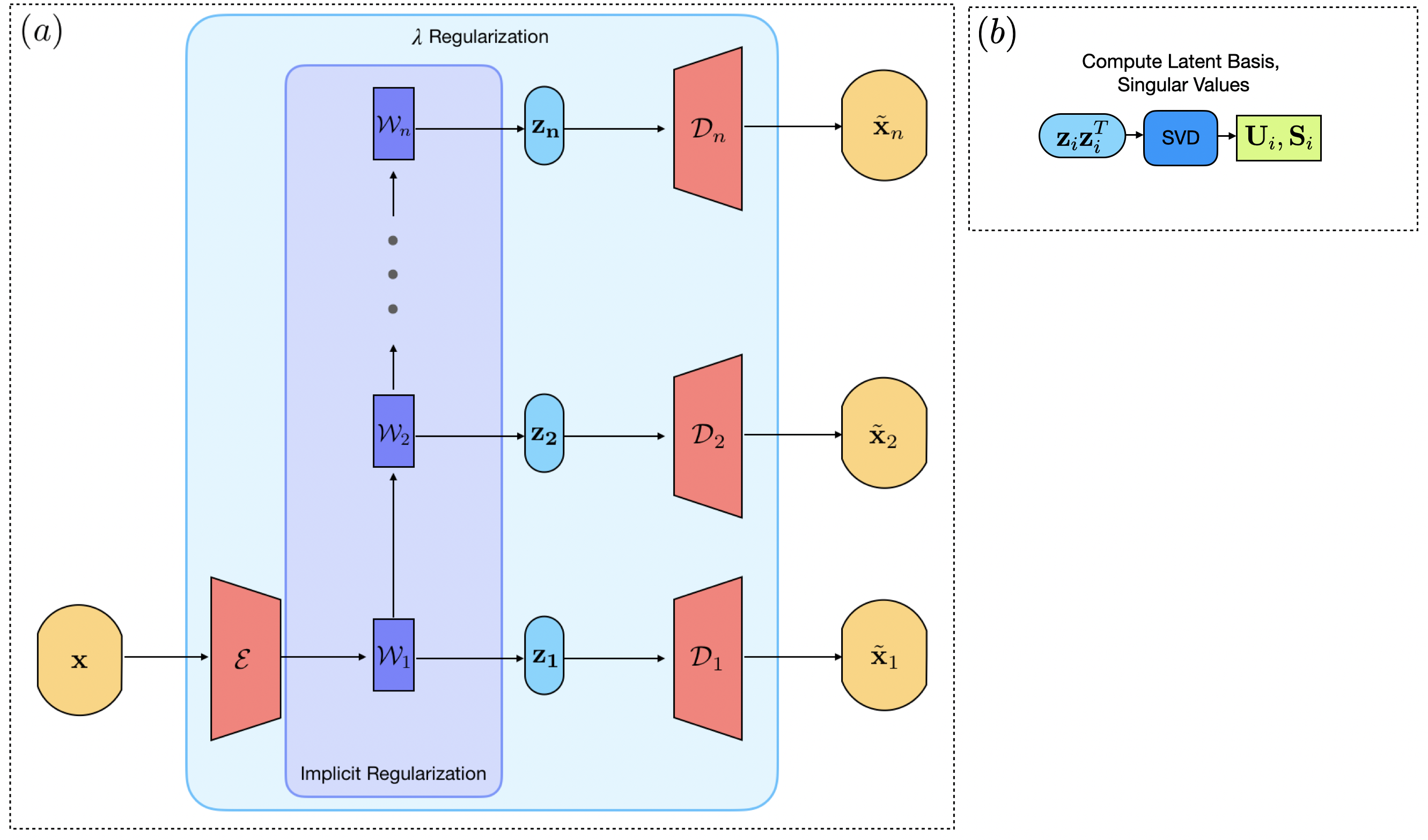}
	\caption{Implicit Rank Minimizing autoencoder with weight decay and branching (IRMAE-WD-B) framework. (a) Network architecture with regularization mechanisms. Note the different paths between the encoder, $\mathcal{E}$, and decoders, $\mathcal{D}_i$. (b) Singular value decomposition of the covariance of the learned latent data representation $\textbf{z}_i$.}
	\label{IRMAE_branch}
\end{figure}

A schematic representation of the IRMAE-WD-B framework is shown in Figure \ref{IRMAE_branch}.
\sout{The encoder, denoted by $\mathcal{E}\left(\boldsymbol{x} ; \theta_{\mathcal{E}}\right)$ reduces the dimension from $N$\MDG{is it $N$ or $d_N$?} to $d_z$. Subsequently,}  
A set of linear networks $\mathcal{W}_i\left(\cdot ; \theta_{\mathcal{W}}\right)$ is included between the encoder and the various decoders $\mathcal{D}_i$ \CRCA{(i.e., $\boldsymbol{z} = \mathcal{W}_i\mathcal{W}_{i-1}\ldots \mathcal{W}_1(\mathcal{E}(\boldsymbol{x};\theta_\mathcal{E});\theta_\mathcal{W})$}).
These decoders  map  the low-dimensional representation back to the full space (e.g., $\tilde{\boldsymbol{x}}_{i}=\mathcal{D}_i\left(\mathcal{W}_i\mathcal{W}_{i-1}\ldots \mathcal{W}_1(\mathcal{E}(\boldsymbol{x};\theta_\mathcal{E});\theta_\mathcal{W})
; \theta_{\mathcal{D}_i}\right)$). Therefore, the loss function for this architecture is represented by
\begin{equation}
\mathcal{L}_i\left(\tilde{\boldsymbol{x}} ; \theta_{\mathcal{E}}, \theta_{\mathcal{W}}, \theta_{\mathcal{D}_i}\right)=\left\langle\left\|\boldsymbol{x}
-\mathcal{D}_i\left(\mathcal{W}_i\mathcal{W}_{i-1}\ldots \mathcal{W}_1\left(\mathcal{E}\left(\boldsymbol{x} ; \theta_{\mathcal{E}}\right) ; \theta_{\mathcal{W}}\right) ; \theta_{\mathcal{D}_i}\right)\right\|_{2}^{2}\right\rangle+\frac{\lambda}{2}\|\theta\|_{2}^{2}.
\label{loss_irmae}
\end{equation}
where $\langle\cdot\rangle$ is the average over a training batch, $\theta_\mathcal{E}$ the weights of the encoder, $\theta_{\mathcal{D}_i}$ the weights of the decoder $i$, and $\theta_\mathcal{W}$ the weights of the linear network which contains the linear layers used in the branch.  We also include a $L_2$ (``weight decay") regularization to the weights with prefactor $\lambda$. In this work $\lambda$ is fixed to $10^{-6}$, following the work from \citep{irmae}.

We note that this architecture will produce a number of loss values corresponding to the number of linear layers. To train the network, we select a branch and backpropagate through it to update the weights. \CRCA{The branch selection process involves randomly selecting a path for each epoch of training, resulting in an improvement in the autoencoder performance. }

To find the low-dimensional representation, we optimize the loss function described in equation \ref{loss_irmae} using an Adam optimiser in pythorch \citep{pythorch}.
We allow the model to choose among 10 different paths. Each path is characterized by an increasing number of linear layers of size 
$d_z\times d_z$ (e.g., path 1 has one linear layer, while path 10 has ten linear layers).
We ran for 1000 epochs, and to further improve the model at the end of the training, we used the path with the lowest mean squared error (MSE) during the last $10\%$ of the total number of epochs.

Post-training, a singular value decomposition (SVD) is applied to the covariance matrix of the latent data matrix  $\textbf{z}$
yielding matrices of singular vectors $\mathsfbi{U}$, and  singular values $\mathsfbi{S}$. 
Then, we can project $\textbf{z}$ onto $\mathsfbi{U}^T$ to obtain $\mathsfbi{U}^T \textbf{z}= \boldsymbol{h}^+ \in\mathbb{R}^{d_z}$ in which each coordinate of $\textbf{h}^+$ is orthogonal and ordered by contribution (here, $\boldsymbol{h}^+$ refers to the projection of latent variables onto manifold coordinates).
This framework reveals the manifold dimension $d_h$ as the number of significant singular values, indicating that a coordinate representation exists in which the data spans $d_h$ directions.
Thus, the encoded data avoids spanning directions  associated with nearly zero singular values (i.e., $\mathsfbi{U} \mathsfbi{U}^T {\bf z} \approx  \mathsfbi{\hat{U}} \mathsfbi{\hat{U}}^T {\bf z} $, where $\mathsfbi{\hat{U}}$ are the singular vectors truncated corresponding to singular values that are not nearly zero). Leveraging this insight, we  extract a minimal, orthogonal coordinate system by projecting   ${\bf z}$ onto $\mathsfbi{\hat{U}}$, resulting in a minimal representation $\mathsfbi{\hat{U}}^T {\bf z} = {\boldsymbol{h}} \in \mathbb{R}^{d_h}$.


\begin{table}
\caption{
Here we list the architecture and parameters utilized in the studies of this paper.  Between the $\mathcal{E}$ and $\mathcal{D}$, there are  $n$ sequential linear layers of shape $d_z \times d_z$. \CRCA{The superscripts in $\boldsymbol{\Psi }$ and $\boldsymbol{g}$ denote   the number of sequential layers with the specified number of node units.} 
}
\centering
\begin{tabular}{ll*{6}{c}r}
\hline
\hline
Case&Function & Shape & Activation  & Learning Rate \\
\hline
\hline
\text{ MFE } & $\boldsymbol{\Psi }$ & $ 9/(200)^4/\Psi + 9 \quad $ & $(\text{elu})^4/\text{elu}$ & $[10^{-4},10^{-5},10^{-6}]$ \\
\hline
\text{ Plane Couette Flow} & $\boldsymbol{\chi}$		& 		502/1004/256/$d_z$ \quad           & ReLU/ReLU/lin         & $[10^{-3},10^{-4}]$ \\
 & $\check{\boldsymbol{\chi}}$		& $d_z$/256/1004/502 \quad           & ReLU/ReLU/lin         & $[10^{-3},10^{-4}]$ \\
& $\boldsymbol{\Psi }$ & $d_h/(200)^6/\Psi + d_h \quad$ & $(\text{elu})^6/\text{elu}$ & $[10^{-4}, 10^{-5}, 10^{-6}]$ \\
& $\boldsymbol{g}$ & $d_h/(500)^2/ d_h \quad$ & $(\text{ReLU})^2/\text{lin}$ & $[10^{-3}, 10^{-4}]$ \\
\hline
\label{table}
\end{tabular}
\end{table}

\subsection{function space modelling: approximation of the Koopman operator \label{framework_function}}

We apply a function space approach to learn an evolution equation either in the low-dimensional representation for plane Couette Flow or full space for the MFE model. We note that the nomenclature used in this section is based on using the low-dimensional representation.
The function space approach  is based on the  infinite-dimensional linear Koopman operator $\Koop$, which describes the evolution of an arbitrary observable $\mathcal{G} (\boldsymbol{h})$ from time $t$ to time $t+\delta t$: $\mathcal{G}(\boldsymbol{h}(t+\delta t))=\Koop\mathcal{G}(\boldsymbol{h}(t))$ \citep{koopman1931,Lasota, Mezic_2023}. 
The tradeoff for gaining linearity is that $\Koop_{\delta t}$ is also infinite-dimensional, requiring for implementation of some finite-dimensional truncation of the space of observables.

Here we use a variant of the EDMD-DL approach \CRCA{ to determine both the dictionary of observables and the corresponding approximation of the Koopman operator, } described in detail in  \citet{edmd_dl_ad}. Given a vector of observables 
$\boldsymbol{\Psi}(\boldsymbol{h}(t))  \CRCA{\in \mathbb{R}^{d_K}}$ \CRCA{(each element of the observable $\boldsymbol{\Psi}$ is a neural network except for the first $d_h$ elements, which correspond to the manifold coordinates themselves)}, now there is a matrix-valued approximate Koopman operator $\Kapp$ such that the evolution of observables is approximated by 
\begin{equation}
\boldsymbol{\Psi}(\boldsymbol{h}(t+\delta t))=\Kapp\boldsymbol{\Psi}(\boldsymbol{h}(t)).\label{eq:ObservableEvolution} 
\end{equation}
Given a time series of observables assembled into a matrix whose columns are the vector of observables at different times,
\begin{equation}
\boldsymbol{\psi}(t)=\left [ \boldsymbol{\Psi}(\boldsymbol{h}(t_1)),~ \boldsymbol{\Psi}(\boldsymbol{h}(t_2)) \ldots \right ]
\end{equation}
and its corresponding matrix at  $ t + \delta t$,
\begin{equation}
\boldsymbol{\psi}(t+\delta t)=\left [ \boldsymbol{\Psi}(\boldsymbol{h}(t_1 +\delta t)),~  \boldsymbol{\Psi}(\boldsymbol{h}(t_2 +\delta t)) \ldots \right ],
\label{obs_matrix}
\end{equation} 
the approximate matrix-valued Koopman operator is defined as the least-squares solution 
\begin{equation}
\Kapp= \boldsymbol{\psi}({t + \delta t }) \boldsymbol{\psi}(t)^\dagger, 
\end{equation}
where $\dagger$ superscript denotes the  Moore-Penrose pseudoinverse. We aim to minimise
\CRCA{the difference between the observable prediction $\tilde{\boldsymbol{\psi}} (\boldsymbol{h}, t+\delta t; \theta)$  (e.g., $\tilde{\boldsymbol{\psi}} (\boldsymbol{h}, t+\delta t; \theta)=  \Kapp \boldsymbol{\psi} (\boldsymbol{h}; \theta)$), and the corresponding observable matrix at $ t + \delta t$ (equation \ref{obs_matrix}). Then, the loss function is defined by}
\begin{equation}
\begin{tabular}{cc}
$\mathcal{L}_K(\boldsymbol{h},\theta_K)=\left \|   \boldsymbol{\psi} (\boldsymbol{h}, t+\delta t; \theta_K)  -  \tilde{\boldsymbol{\psi}} (\boldsymbol{h}, t+\delta t; \theta_K)    \right \|_F$ =\\
$= \left \|   \boldsymbol{\psi}_{\text{NN}} (\boldsymbol{h}, t+\delta t; \theta_K)  -  \boldsymbol{\psi}_K (\boldsymbol{h}, t+\delta t; \theta_K) \boldsymbol{\psi}_K(\boldsymbol{h}, t; \theta_K)^+  \boldsymbol{\psi}_{\text{NN}} (\boldsymbol{h}, t; \theta_K)  \right \|_F$
\end{tabular}
\label{loss_batch}
\end{equation}
\CRCA{where $\left\| \cdot \right\|_F$ refers to the Frobenius norm.}
 Here, 
 $\theta_K$ stands for  the weights of the neural networks. We take advantage of automatic differentiation, and we compute the gradient of $\partial\mathcal{L}_K/\partial \theta_K$ directly, enabling to
find $\Kapp$ and the set of observables $\boldsymbol{\Psi}(\boldsymbol{h})$ simultaneously using the Adam optimizer \citep{Adam}.  
For more details, we refer  to \cite{edmd_dl_ad}, in which this Koopman methodology is extensively explained. 



\MDGrevise{For time evolution predictions, a `pure' Koopman approach would simply apply Eq.~\ref{eq:ObservableEvolution}. (In \cite{edmd_dl_ad}, this method was denoted KDL$_\mathrm{oo}$).   While respecting the linearity of the exact infinite-dimensional formulation, this approach struggles to produce accurate results over long time horizons even for simple systems, not to mention complex chaotic ones \citep{edmd_dl_ad}. Essentially, small errors accumulate and a drift of the dynamics are observed. Nevertheless, it has been found to work  well for  systems featuring steady and oscillatory behaviours -- see \citet{edmd_dl_ad,pipe_flow}. }

\MDGrevise{A slight modification of Eq.~\ref{eq:ObservableEvolution} has been found in past work \citep{Qianxiao,Junker,van2023reprojection,edmd_dl_ad} to generate substantially better time-evolution predictions for simple systems, and we show below that it also works extremely well for complex chaotic systems. In this approach, at each time step the state $\bh(t)$ is used to construct $\boldsymbol{\Psi}(\bh(t))$, which is then evolved according to Eq.~\ref{eq:ObservableEvolution} to yield $\boldsymbol{\Psi}(\bh(t+\delta t))$. Finally $\bh(t+\delta t)$ is obtained by projection -- recall that the first $d_h$ elements of $\boldsymbol{\Psi}$ are simply the state (in manifold coordinates) $\bh$. In summary, one time step of this approach is given as \begin{equation}
\boldsymbol{h}(t+\delta t)=\mathsfbi{P}\mathsfbi{K}\boldsymbol{\Psi}(\boldsymbol{h}(t))
\label{eq:projection}
\end{equation}
where $\mathsfbi{P}$ represents  a projection operator that maps down from the full vector of observables to just the state. We refer to this approach as `projected Koopman dynamics'. (In \cite{edmd_dl_ad}, this method was denoted KDL$_\mathrm{so}$).
Because $\boldsymbol{\Psi}(\bh)$ is nonlinear, this is a nonlinear evolution equation for $\bh$. (It is worth noting that other Koopman-based formalisms for time evolution such as that of \citet{khodkar2019koopman} are also nonlinear.)}

\MDGrevise{Although this approach originated in the space of observables, in the end it is simply a nonlinear discrete-time evolution equation in the state space of $\bh$ with a particular structure. It is a composition of a nonlinear mapping to a high dimensional observable space given by $\boldsymbol{\Psi}$, a linear evolution of $\boldsymbol{\Psi}$ with $\mathsfbi{K}$ according to the Koopman formalism, and projection back from the observable space to the manifold coordinates with $\mathsfbi{P}$. In the deep learning context, it is a nonlinear neural network followed by two linear layers. In combination with dimension reduction onto an invariant manifold, we refer to this framework as `Koopman Data-driven Manifold Dynamics' (KDManD). }

To find an effective architecture for lifting the state space to the observable space, we use a trial-and-error approach, varying the network architecture, the number of dictionary elements, and the activation functions. Table 1 summarises the network parameters that yielded the best results. In this work, we select 200 dictionary elements to lift the state space to the observable space. To optimize the loss function described in equation \ref{loss_batch}, we use an Adam optimiser in pythorch \citep{pythorch}.  We select a batch size of 5000, and we train the models for 5000 epochs using a learning rate scheduler that systematically decreases its value by three orders of magnitude, from $10^{-4}$ to $10^{-6}$ over an equivalent span of three epochs. At this point, we see no improvement in the reconstruction error. 

\subsection{state space modelling: Neural ODEs \label{framework_node}}
\MDGrevise{We use a stabilized neural ODE framework for state space modeling in the latent space.  Rather than Eq.~\eqref{eq:StateEvolution}, we use a slight modification:
\begin{equation}
\frac{d\bh}{dt}=\bg(\bh)-\mathsfbi{A}\bh.\label{eq:stabNODE}
\end{equation}
Here ${A}_{ij}=-\beta\delta_{ij}\sigma_i(\boldsymbol{h})$, $\sigma_i(\boldsymbol{h})$ is standard deviation of the $i$th component of $\boldsymbol{h}$, $\beta$ is a fixed parameter and $\delta_{ij}$ is the Kronecker delta. This modification, with a small value of $\beta$ it has been found to stabilize the dynamics against spurious growth of fluctuations without compromising accuracy of predictions \cite{alec_chaos,alec_node,Floryan.2024}.}
\MDGrevise{Representing the vector field $\boldsymbol{g}$ on the manifold as a neural network with weights $\theta_f$, we can time-integrate Eq.~\ref{eq:stabNODE}}  between $t$ and $t+\delta t$ to yield a prediction $\tilde{\boldsymbol{h}}(t+\delta t)$, such as
\begin{equation}
\tilde{{\boldsymbol{h}}}(t+\delta t)=\boldsymbol{h}(t)+\int_{t}^{t+\delta t}\left(\boldsymbol{g}(\boldsymbol{h}(t');\theta_f) + \mathsfbi{A}\boldsymbol{h}(t')\right)dt'.   
\end{equation}
Given data for $\boldsymbol{h}(t)$ and $\boldsymbol{h}(t +\delta t)$ for a long time series we can train $\boldsymbol{g}$  to minimize the $L_2$ difference between the prediction $\tilde{\boldsymbol{h}}(t+\delta t)$   and the known data  $\boldsymbol{h}(t+\delta t)$. Then the loss function is defined by
\begin{equation}
\mathcal{L}_g=\frac{1}{dN}\sum_{i=1}^N||\boldsymbol{h}(t_i+\delta t)-\tilde{\boldsymbol{h}}(t_i+\delta t)||_2^2.
\label{loss_node}
\end{equation}
We use automatic differentiation to determine the derivatives of $\boldsymbol{g}$ with respect to  $\theta_f$.    To optimize the loss function described in equation \ref{loss_node}, we use an Adam optimiser in pythorch \citep{pythorch}.

\section{Results \label{Results}}


\subsection{MFE model}

\subsubsection{Description of MFE data \label{mfe_model}}

The MFE model, developed by \citet{moehlis2004low}, is a low-dimensional model that describes the dynamics of an incompressible fluid between infinite, free-slip walls driven by a spatially sinusoidal body force. The domain is periodic of size $L_x \times L_y \times L_z$, with infinite, parallel walls at $y = \pm L_y/2$  and periodic boundaries 
in $x$ and $z$.
Here, $x$, $y$, and $z$ are the streamwise, wall-normal, and spanwise coordinates, respectively.
The model displays features consistent with turbulence in the transition regime, with long periods of turbulent behavior with infrequent quasi-laminarization events (also called hibernating \citep{Xi2010} or quiescent \citep{Hamilton1995} intervals) and ultimately full laminarization at long times \citep{Hamilton1995, Xi2010, Hof2006}. 
The model is an approximation to the Navier-Stokes equations, generated through a severe truncation of the Fourier Galerkin method, that represents the flow as combinations of spatial Fourier modes $\boldsymbol{v}_i (\boldsymbol{x})$.
The velocity field $\boldsymbol{v}$ is given by a superposition of nine models 
\begin{equation}
\boldsymbol{v} (\mathbf{x},t) = \sum\limits^9_{i=1} a_i(t) \boldsymbol{v}_i(\mathbf{x})
\end{equation}
with modes describing the laminar flow profile and turbulent deviations caused by streaks, vortices, and the interactions between them.
The dynamics of the system are captured through the evolution of the nine modal amplitudes $a_i(t)$, governed by nine ordinary differential equations given by the Galerkin projection, which can be found explicitly in \citet{moehlis2004low}.  For this case, we perform no dimension reduction, so $\bx=\bh=[a_1, a_2,\ldots,a_9]^T$.

Domain dimensions of $L_x=4\pi$, $L_y=2, L_z=2\pi$, with a channel Reynolds number of $Re=400$, have been shown to produce turbulent flow dynamics with sufficient self-sustainment to be captured through data-driven modeling \citep{Srinivasan2019}.
\CRCA{We generated 100 unique time series as training data by integrating the MFE equation using a fourth-order Runge-Kutta method with a time step of 0.5 (i.e., $\delta t=0.5$). Each time series captures the transient turbulent state, featuring intervals of turbulence interrupted by quasi-laminarization events, ultimately leading to full laminarization at long  time. The flow is often characterized using the total kinetic energy (KE), defined as \( KE = \frac{1}{2} \sum_{i=1}^{9} a_i^2 \). In this context, the turbulent state corresponds to low energy, while the laminar state is high energy. \MDGrevise{After a long turbulent transient,} every time series eventually evolves to the  laminar fixed point \( a_i = \delta_{i1} \).
The time series were generated with initial conditions for eight of the amplitudes set as follows: \( (a_1, a_2, a_3, a_5, a_6, a_7, a_8, a_9) = (1, 0.07066, -0.07076, 0, 0, 0, 0, 0) \). The initial value of \( a_4 \) was randomly chosen within the range \( a_4 \in [-0.1, 0.1]\). These initial conditions have been shown to produce chaotic dynamical data with quasi-laminarization events \citep{Eivazi}. The results are presented with time normalized using the Lyapunov time, \( \tau_L \), for the system (e.g.,  \( \tau_L \approx 61 \)  according to \citet{Racca2022}). The first 1000 time steps of each time series were discarded to eliminate dependence on initial conditions. Each time series was evolved until the laminar state was reached, requiring a variable number of time steps depending on the initial conditions. The resulting training data set comprised approximately 2 million snapshots of MFE amplitudes, with a mean lifetime of \( 164\tau_L \). An example of the evolution of three of the amplitudes from a trajectory, as well as the total kinetic energy of the trajectory, $E_k = \frac{1}{2} \sum_1^9 a_i^2$, is shown in Figure \ref{intro}a. }


\subsubsection{Koopman  MFE model  }

\begin{figure}
\centering
\begin{tabular}{cc}
\includegraphics[width=0.4\textwidth]{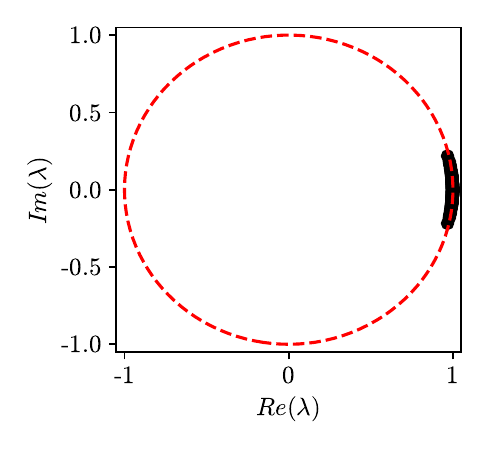}&
 \includegraphics[width=0.52\textwidth]{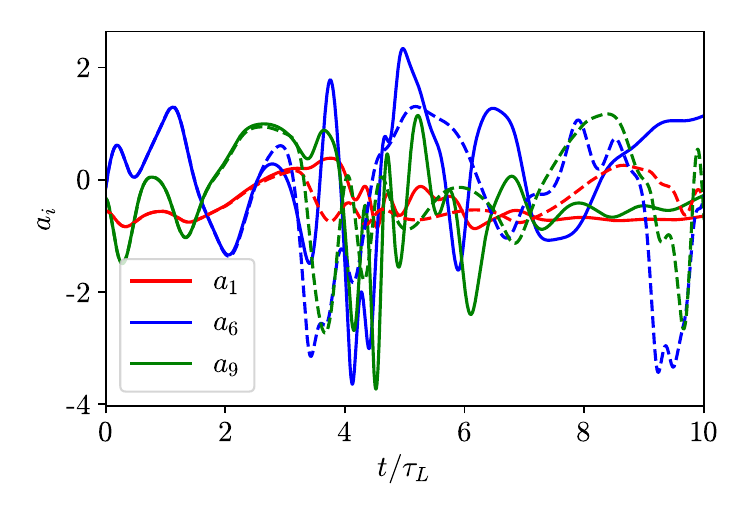}  \\(a) & (b)\\
\end{tabular}
\caption{(a) Eigenvalues of the Koopman operator for MFE model
\CRCA{(b) Temporal evolution of three amplitudes $a_1$, $a_6$ and $a_9$ for the true MFE solution (solid lines) and Koopman predictions (dashed lines) for the same initial condition.}
}
\label{VelocityStatistics}
\end{figure}

First, we apply Projected Koopman Dynamics to the MFE model described in section \ref{mfe_model} with $\delta t=0.5$. 
Figure  \ref{VelocityStatistics}a shows the eigenvalues of the Koopman operator with 200 dictionary elements.
\CRCA{We observe eigenvalues within the unit circle (i.e., $|\lambda_k|\approx 0.99$).
If we use the linear evolution described by equation
\ref{eq:ObservableEvolution}, given that the $|\lambda_k|< 1$, the dynamics will decay. But, by using the projection, described by Eq.~\ref{eq:projection},  we achieve a good approximation of the true dynamics. }
%
%
\CRCA{Figure \ref{VelocityStatistics}b shows the temporal evolution of three amplitudes $a_1$, $a_6$ and $a_9$ for the true MFE solution  and Koopman predictions  for the same initial condition up to $10\tau_L$. The predictions show an exceptional agreement with the true solution up to 3$\tau_L$, then the dynamics diverge due to their chaotic nature. }

The success of our Koopman-based model is first assessed in its ability to reconstruct the time-averaged velocity statistics of the turbulent regime of the MFE model, demonstrating a base level of competence vital for any data-driven modeling technique.
Using 100 randomized initial conditions, we used our Koopman model to forecast the evolution of the modal amplitudes over a period of 100 $\tau_L$.
These same initial conditions were similarly evolved using the MFE equations over the same time period. 
The true and predicted amplitudes were then projected onto the corresponding spatial modes, and the time-averaged velocity statistics of the corresponding model were calculated.

\begin{figure}
\centering
\begin{tabular}{cc}
\includegraphics[width=0.59\textwidth]{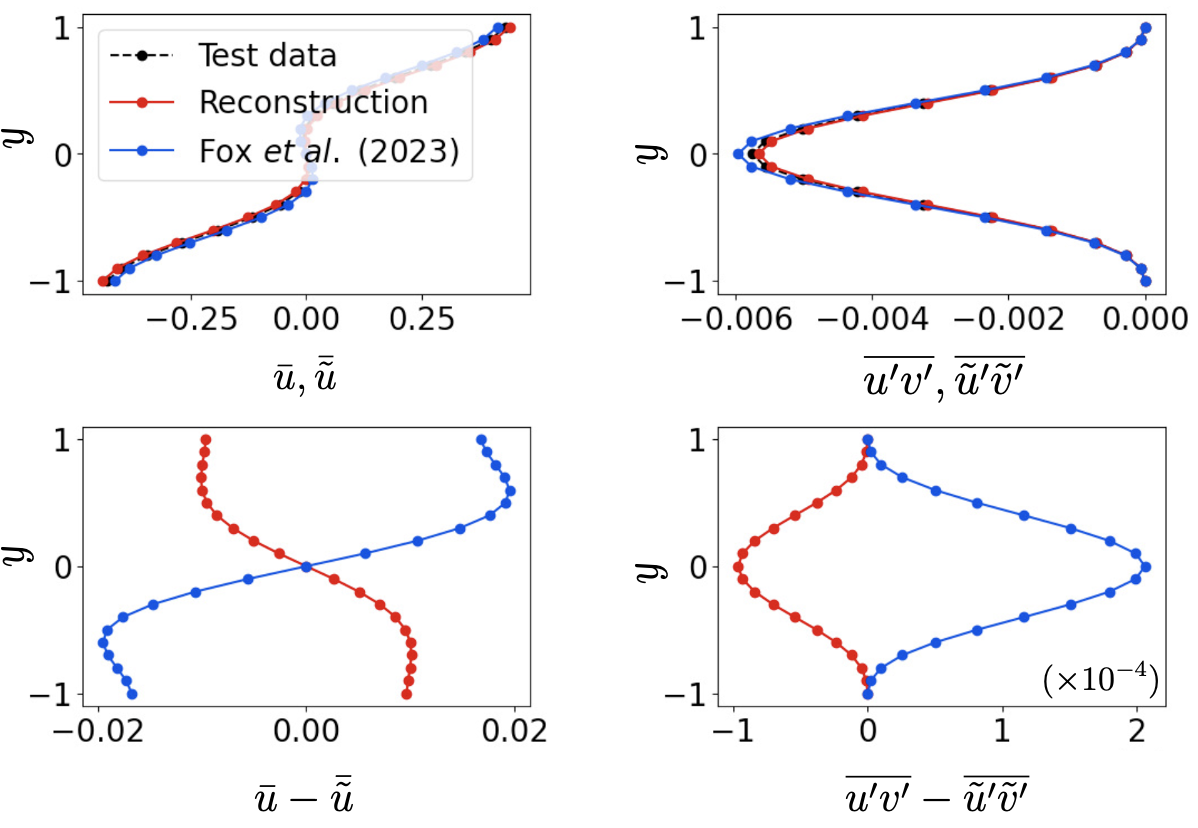}&
\includegraphics[width=0.41\textwidth]{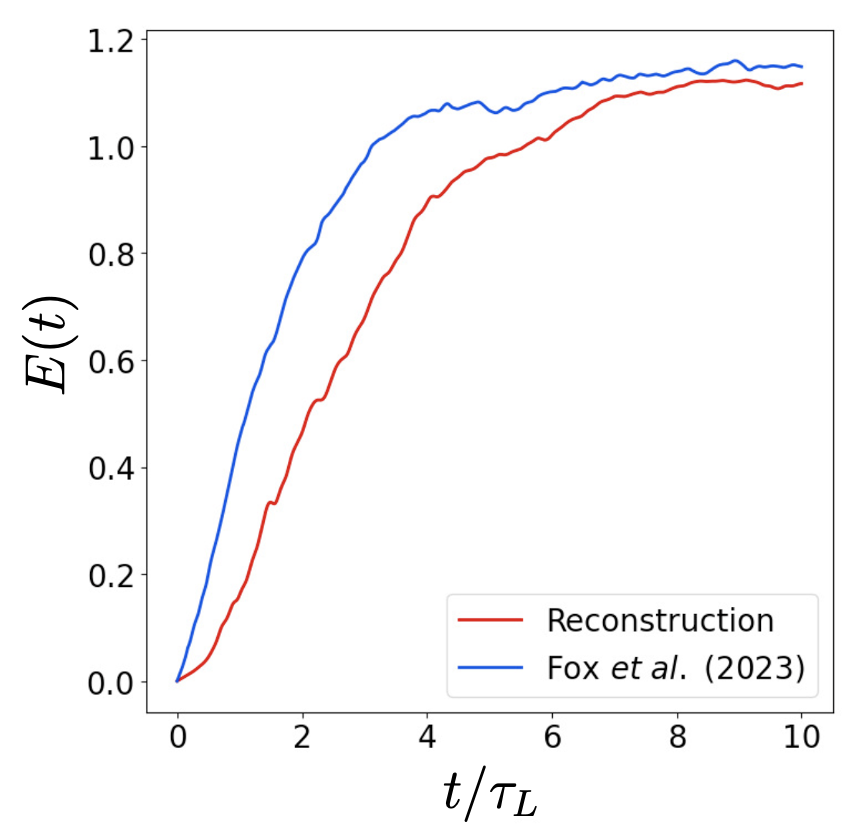}\\
(a) & (b)\\    
\end{tabular}
\caption{(a) The top left and right panels display the mean streamwise velocity and Reynolds shear stresses for the full field of both the test and training data, respectively. The bottom panels show the differences between the true values and the predicted values. Results from \citet{Fox2023} are also included. (b) Ensemble averaged short-time error tracking of the reconstruction of the Koopman model, with comparison to prior study by \citet{Fox2023}.
}
\label{ShortTimeError}
\end{figure}

Figure \ref{ShortTimeError}a shows the mean streamwise velocity and Reynolds shear stress from the resulting true and reconstructed flow fields, as well as the difference between these two measurements. The Koopman-based model captures the structure and magnitude of the velocity statistics extremely well, with the deviation from the true values remaining comparatively small throughout the spatial domain.
The Koopman model outperforms the model 
from \citet{Fox2023}, reducing the error in the forecasts of both the velocity profile and the Reynolds shear stress. 

The Koopman-based model was next assessed on its ability to precisely forecast the short-time evolution of the modal amplitudes in the turbulent regime. 
Using 1000 random and unique initial conditions, we forecasted the evolution of the MFE modes using our Koopman model for a time period of 10 $\tau_L$, with the same initial conditions evolved separately by the MFE equations over the same period.
The accuracy of our forecasts was then evaluated by calculating a time-dependent, \CRCA{relative ensemble-averaged error over the testing time period $E(t) = \left< ||a(t) - \tilde{a}(t)||_2 \right>/\left< ||a(t) ||_2 \right> $} (see figure \ref{ShortTimeError}b).
The Koopman-based model performs very well, accurately forecasting the evolution of the modal amplitudes for nearly  one $\tau_L$ with minimal error.
The ensemble-averaged error of the forecasts grows very slowly over the testing period, requiring several Lyapunov times to completely deviate from the true amplitude evolution, further demonstrating the short-time accuracy of the Koopman-based model.
The Koopman model again outperforms the results of \citet{Fox2023}, with the new data-driven model reducing the error growth significantly in the testing window.

\begin{figure}
\centering
\includegraphics[width=.9\textwidth]{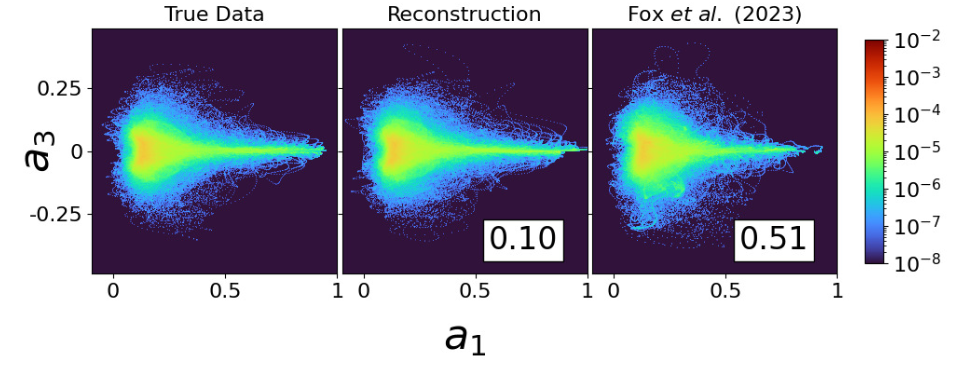}
\caption{ Joint probability density function (jPDF) of $a_1$ and $a_3$ for the true data,  the Koopman model, and the work by \citet{Fox2023}, corresponding to left-to-right panels, respectively. For the models, we have added the corresponding value of the 
relative mean squared error (rMSE).
Note the logarithmic scale.}
\label{JointPDF}
\end{figure}

We next determined the ability of the Koopman-based model to predict the long-time structure and occurrence of quasi-laminarization events in turbulent regimes.
In the MFE model, quasi-lamizarization is characterized by an increase in the first modal amplitude, associated with the mean shear flow profile, and a decrease in the amplitude of the remaining eight modes, which correspond with the turbulent fluctuations. After a conditionally-dependent time, these values collapse back towards their original range as the flow evolves back into the turbulent regime.
To observe the frequency and structure of these events, we generated a test data set containing 100 trajectories evolved with the MFE model for 100 Lyapunov times, separately evolving the initial conditions of these trajectories with the Koopman-based model; as we are here interested solely in the turbulent behavior, all full re-laminarizations were removed from both data sets.
Shown in Figure \ref{JointPDF} is the joint probability density function (PDF) of $a_1$ and $a_3$ for the true and reconstructed trajectories, where the quasi-laminarization events can be seen in the narrowing tail on the right side of the plots.
The reconstruction by the Koopman mode closely matches that of the true data, adequately capturing the turbulent and quasi-laminar states.
\CRCA{The accuracy is further quantified by the relative mean squared error, $rMSE = MSE(PDF_{true} - PDF_{pred}) / \mathcal{P} - 1$, calculated by taking the mean squared error (MSE) between the two joint PDFs and normalizing it by the MSE between two joint PDFs generated from true
trajectories of the same length with different initial conditions, $\mathcal{P}$, and subtracting one; here, the relative MSE is 0.1,} shown inset in the middle panel of Figure \ref{JointPDF}, a very close match and an improvement over a previous study \MDGrevise{using the state-space based CANDyMan approach} \citep{Fox2023}.

\begin{figure}
\centering
\includegraphics[width=1\textwidth]{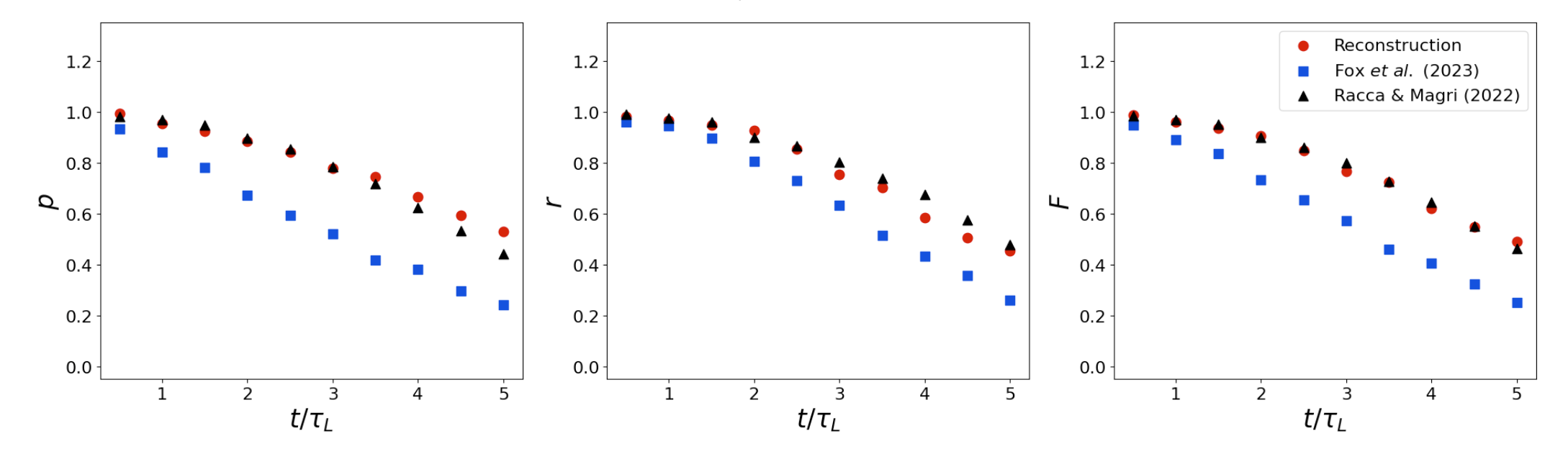}
\caption{Precision, recall, and \textit{F-score} of extreme event forecasting of the MFE model by the Koopman model, with comparison to prior studies by \citet{Fox2023} and \citet{Racca2022}.}
\label{F-score}
\end{figure}

Finally, we assessed the ability of the data-driven model to forecast the occurrence of a quasi-laminarization event, here defined as an increase in the kinetic energy of the system such that $E_k 
>0.1$.
Using 1000 initial conditions, we used the MFE equations and the Koopman model to evolve the modal amplitudes separately for a period of 5.5 $\tau_L$.
We then segmented the trajectories into 0.5 $\tau_L$ time windows, where the window at time $\tilde{t} = t / \tau_L$ refers to the time between $\tilde{t}$ and $\tilde{t} + 0.5 \tau_L$.
Each window was then investigated for a quasi-laminarization event in both the true and forecasted data sets for each trajectory, and the two observations were compared.
If an extreme event occurred in both the true and reconstructed trajectories, this was labeled as a \textit{true positive} ($TP$).
If the true trajectory exhibited an extreme event, but the Koopman model failed to predict one, this was labeled as a \textit{false negative} ($FN$).
If the model forecasted an extreme event when the true trajectory showed none, it was identified as a \textit{false positive} ($FP$) \citep{Racca2022}.
The total number of each identification type in each window was tabulated and used to calculate the precision $p$, recall $r$, and \textit{F-score} $F$, where $p = TP / (TP+FP)$, $r = TP / (TP+FN)$, and $F = 2 / (p^{-1} + r^{-1})$.
Therefore, $p=1$ indicates that all predicted events did occur, $r=1$ indicates that all events were predicted, and $F=1$ indicates that both are true.
Figure \ref{F-score} displays   the value of each metric at each time window over the testing period, as well as a comparison to the results of \citet{Racca2022}, which used an echo state network to generate a data-driven model of the system, as well as to our previous study \citet{Fox2023}.
The Koopman model performs very well, bettering the forecasting ability of the model by \citet{Fox2023}, and equaling the forecasting ability of \citet{Racca2022} at all time windows across the testing period.
Remarkably, the Koopman model can achieve this despite being Markovian, while the model of \citet{Racca2022} is notably non-Markovian (as they used an echo state network). 

\CRCA{Finally, when evaluating long-term full laminarization, the Koopman models do not transition to the laminar state, instead maintaining the dynamics in the turbulent state. We note that the initial conditions of the  training dataset described in section \ref{mfe_model}  are far from the laminar state, and subsequently, only a small portion of the data reflects the dynamics in the laminar state. To address this issue, a new training dataset with initial conditions closer to the laminar state should be included during training of the Koopman model.
We note that \citet{pan2024lifting} discussed the challenges  of applying Koopman methods to systems with multiple attractors, which will also apply to systems with long chaotic transients (`chaotic saddles').
}

\subsection{Plane Couette Flow}

\subsubsection{Description of PCF data}

We perform DNS of an incompressible viscous fluid  inside of two plates moving in opposite directions at the same speed. A schematic representation of the numerical setup is shown in Figure \ref{intro}b. The dimensionless form of the continuity and momentum equations are respectively expressed as

\begin{equation}\label{div}
 \nabla \cdot \boldsymbol{v}=0,
\end{equation}
\begin{equation*}
\frac{\partial \boldsymbol{v}}{\partial t}+
\boldsymbol{v} \cdot\nabla \boldsymbol{v} 
= -\nabla p
+ Re^{-1}  \nabla^2 \boldsymbol{v},
\label{NSE}
\end{equation*}
here $t$, $\boldsymbol{v}=[v_x,v_y,v_z]$ and $p$ stand for the time, the velocity and the pressure, respectively. Here, $Re$ is the Reynolds number, defined as $Re = Ud/\mu$, where $U,d$  and $\mu$ stand for the plate velocity, the distance between plates and  the kinematic viscosity, respectively. We consider a minimal flow unit (MFU) of the PCF system, and the 
 computational cell follows the domain presented by \citet{Hamilton1995} with $Re=400$ and a domain size $[L_x, L_y, L_z] = [1.75\pi, 2, 1.2\pi]$; where, $x\in[0,L_x]$, $y\in[-1,1]$ and $z\in[0,L_z]$ for the streamwise, wall-normal and spanwise directions, respectively. 
 The data was generated  using the  pseudo-spectral Channelflow code  on a grid of dimensions $[N_x, N_y, N_z] = [32, 35, 32]$ in the $x$, $y$, and $z$ directions,  respectively \citep{gibson2012channelflow,alec_couette}.
We impose periodic boundary conditions in $x$ and $z$ and no-slip and no-penetration boundary conditions in $y$.

Equation \ref{NSE}, under the boundary conditions described above, is invariant (and its solutions equivariant) under the discrete symmetries of point reflections about $[x, y, z] = [0, 0, 0]$:
\begin{equation}
\mathcal{P}[(v_x, v_y, v_z, p)(x, y, z, t)] = (-v_x, -v_y, -v_z, p)(-x, -y, -z, t),
\end{equation}
reflection about the $z = 0$ plane:
\begin{equation}
\mathcal{R}[(v_x, v_y, v_z, p)(x, y, z, t)] = (v_x, v_y, -v_z, p)(x, y, -z, t),
\end{equation}
and rotation by $\pi$ about the z-axis:
\begin{equation}
\mathcal{RP}[(v_x, v_y, v_z, p)(x, y, z, t)] = (-v_x, -v_y, v_z, p)(-x, -y, z, t).
\end{equation}
In addition to the discrete symmetries, there are also continuous translation symmetries in \(x\) and \(z\). 
\begin{equation}
\mathcal{T}_{\sigma_x, \sigma_z}[(v_x, v_y, v_z, p)(x, y, z, t)] = (v_x, v_y, v_z, p)(x + \sigma_x, y, z + \sigma_z, t).
\end{equation}

We incorporate all these symmetries into the POD representation of the data (as done by \citet{smith2005pod,alec_couette}). Then, we apply the method of slices  to achieve phase alignment of the flow in the spanwise ($z$) direction \citep{budanur_jfm_2017}. This phase alignment in $z$ allows us to fix the position of the low-speed streak in the center of the domain, enhancing the predictive capability of the reduced order models.

Different initial conditions are chosen, and after removing the transient data and the last time to avoid laminar states,
we accumulate a total of 91,562 time units of data \CRCA{which live in the turbulent attractor}, stored at intervals of one time unit (i.e., $\delta t=1$). The data was then divided into $80\%$ for training and $20\%$ for testing purposes.  

\begin{figure}
\centering
\begin{tabular}{cc}
\includegraphics[width=0.58\textwidth]{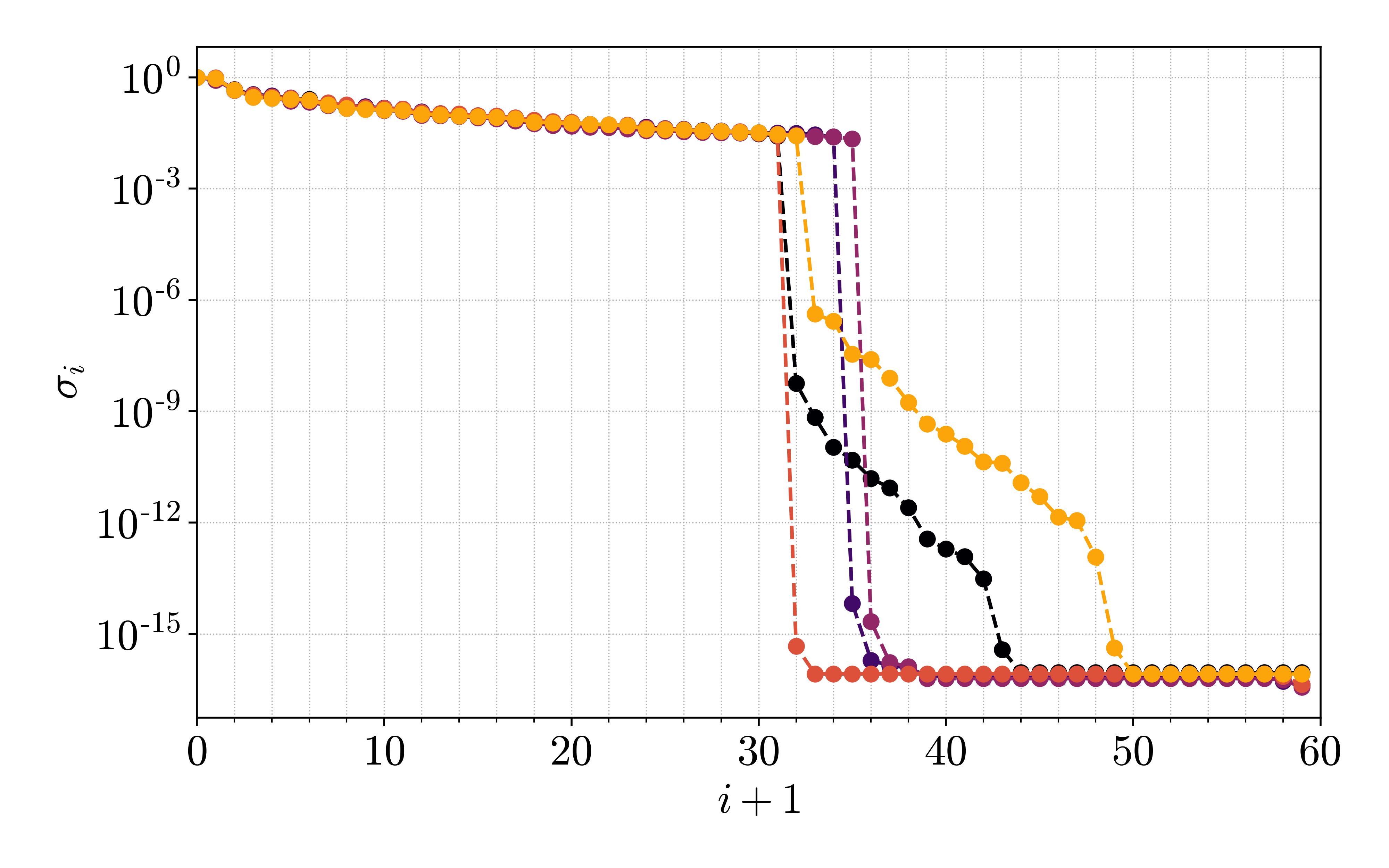}&
\includegraphics[width=0.42\textwidth]{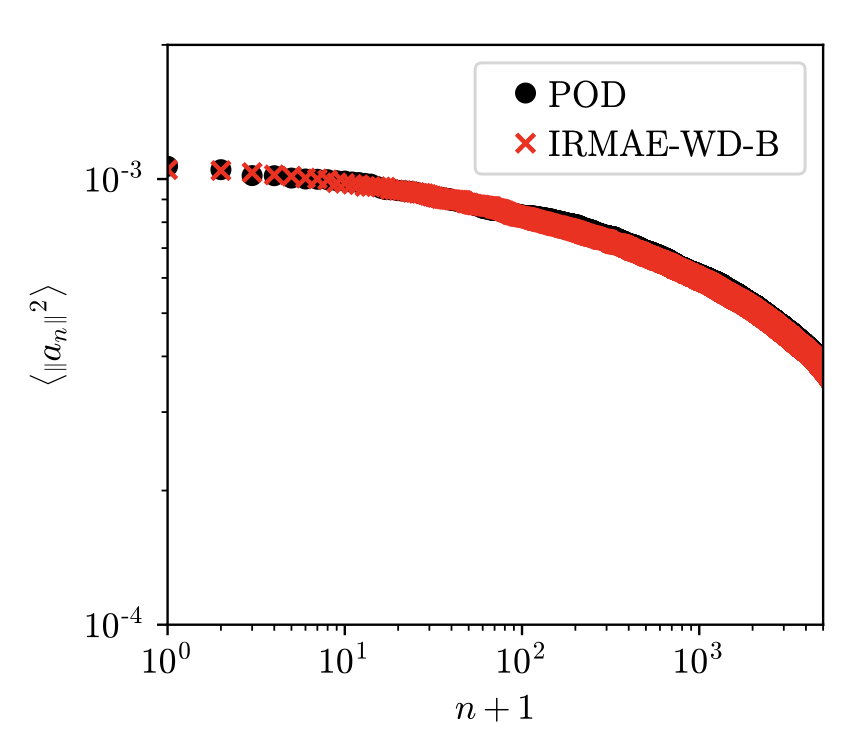}
\\
(a)     & (b)
\end{tabular}
\begin{tabular}{c}
\includegraphics[width=0.6\textwidth]{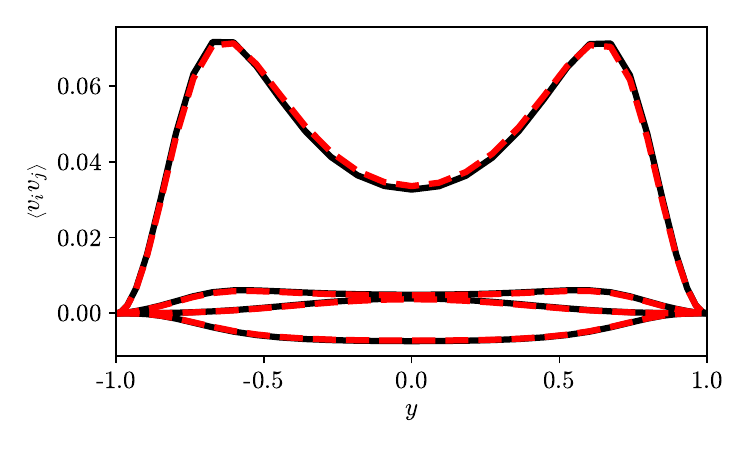}
\\
(c)    \\
\end{tabular}
\caption{Identification of the manifold coordinates. (a) Normalised singular values of the covariance of the latent space data for five different trials. (b) \CRCA{Reconstruction of} $\left \langle \left \| a_n \right \|^2 \right \rangle$ (mean-squared POD coefficient amplitudes) for the test data from 502 POD modes and the IRMAE-WD-B  with $d_h = 36$. (c) Reconstruction of the Reynolds stresses for the DNS and IRMAE-WD-B  with $d_h = 36$.
}
\label{irmae_figure}
\end{figure}

\subsubsection{Finding Manifold Coordinates}

The initial step in building the state space or function space models  involves reducing the dimensionality of the full state ($\mathcal{O}(10^5)$) to its manifold coordinates ($\mathcal{O}(10^1)$). This step is crucial because lifting the state space to a high-dimensional observable space,
would be computationally intractable using the full space.
The first step in building this low-dimensional model is to perform a linear reduction with POD. We select 256 modes, resulting in a vector of POD coefficients $a(t)$, which contains $99.8\%$ of the energy. Projecting onto these modes gives rise to a 502-dimensional system. 
For more details about the POD reduction, we refer  to \citep{alec_couette}, in which the POD framework was  extensively explained for MFU PCF.
After converting the data into POD coefficients and eliminating the low-energy modes, the next step involves training IRMAE-WD-B for nonlinear dimension reduction.

Figure \ref{irmae_figure}a shows the normalized singular values of the \MDGrevise{covariance of the} latent space data for five different trials. It shows that the singular values drop several orders of magnitude in the range of $32 < d_h < 36$, with some  drops reaching magnitudes as low as  $\mathcal{O}(10^{-14})$. The relative error in the POD coefficients across all the trials, measured as $\left \langle \left \| a -\tilde{a} \right \| \right \rangle/  \left \langle \left \| a  \right \| \right \rangle \approx 0.06$. To demonstrate that the latent space spanned  for IRMAE-WD-B is capable of reconstructing accurately the POD coefficients, Figure \ref{irmae_figure}b  shows the $\left \langle \left \| a_n \right \|^2 \right \rangle$ (mean-squared POD coefficient amplitudes) for the test data from 502 POD modes and the IRMAE-WD-B  with $d_h = 36$. The choice of $d_h = 36$ reflects the highest predicted latent dimension among the five trials, representing a conservative choice. Finally,  Figure \ref{irmae_figure}c shows the Reynolds stresses of the test data, and those from the truncated latent dimension at $d_h = 36$, and we observe an exceptional agreement between them.

Across the five independent trials, we observed that IRMAE
-WD-B does not follow a consistent path; instead, it demonstrates an alternation of varying numbers of linear layers across all trials. This variability suggests that a degree of randomness may contribute to improving the dimension estimate.  We note that \citet{alec_couette} found that a latent space with a dimensionality of $d_h=18$ is capable of faithfully capturing the short-time tracking and long-time statistics of the system, including the leading Lyapunov exponents, Reynolds stresses, and Power-Input dissipation. \CRCA{To determine this dimension, they train autoencoders until the MSE plateaus, and then train NODE at different dimensions. They did not observe quantitative improvements beyond $d_h=18$.}

\subsubsection{State and function space  modelling in manifold coordinates \label{couette_Results}}

\begin{figure}
    \centering
    \begin{tabular}{cc}
        \includegraphics[width=0.34\linewidth]{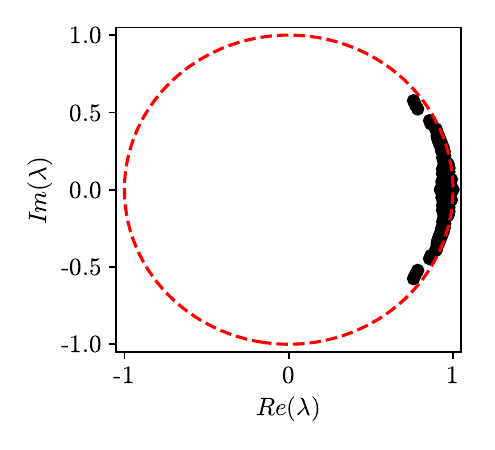} &
        \includegraphics[width=0.57\linewidth]{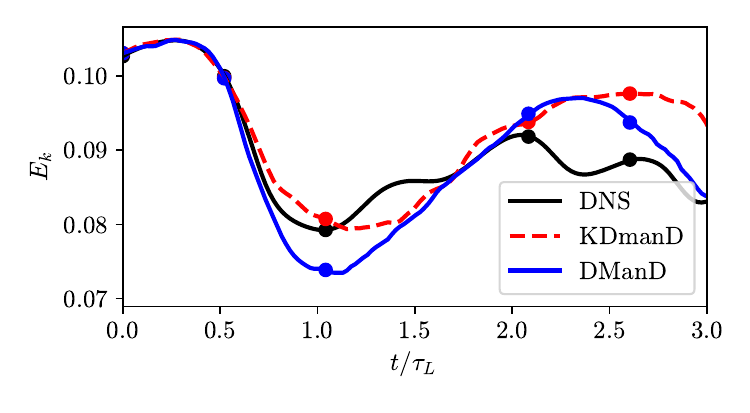} \\
        (a) & (b) \\
    \end{tabular}
    \begin{tabular}{c}
        \includegraphics[width=\linewidth]{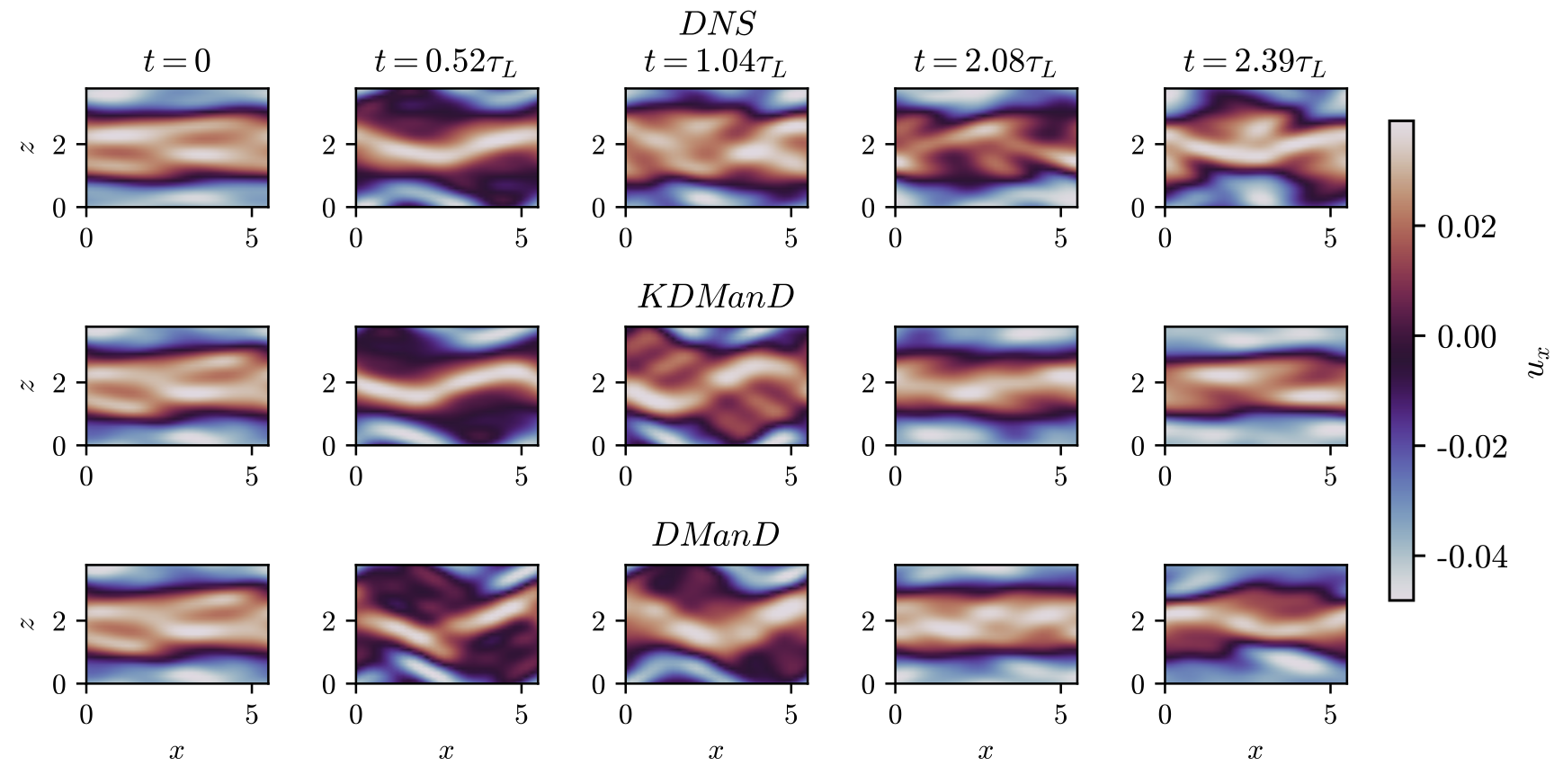} \\
        (c)
    \end{tabular}
    \caption{(a) Eigenvalues of the Koopman operator. (b) Kinetic energy ($E_k$) of the flow for the DNS, KDManD and DManD with $d_h=36$  up to $t/\tau_L=3$. (c) Two-dimensional representation of the dynamics in a $z$-$x$ plane ($y=0$) with $u_x$ for the DNS and the models. Time is displayed at the top of each panel and corresponds to the marked points in panel (b).}
    \label{IC}
\end{figure}

\begin{figure}
    \centering
    \begin{tabular}{cc}
        \includegraphics[width=0.5\linewidth]{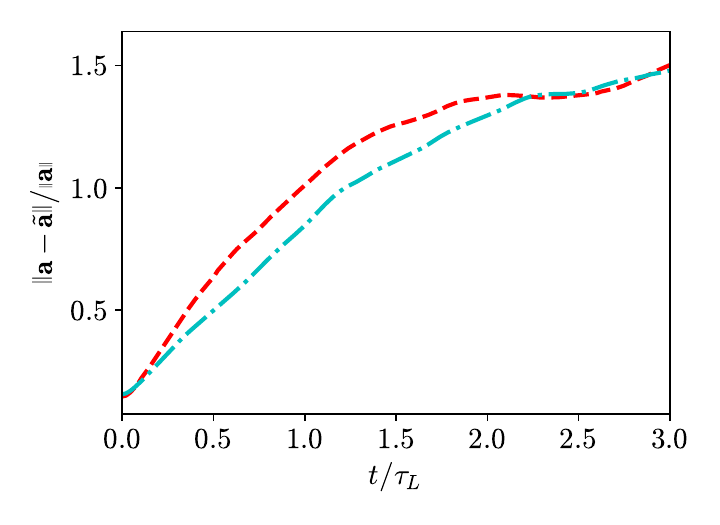} &
        \includegraphics[width=0.5\linewidth]{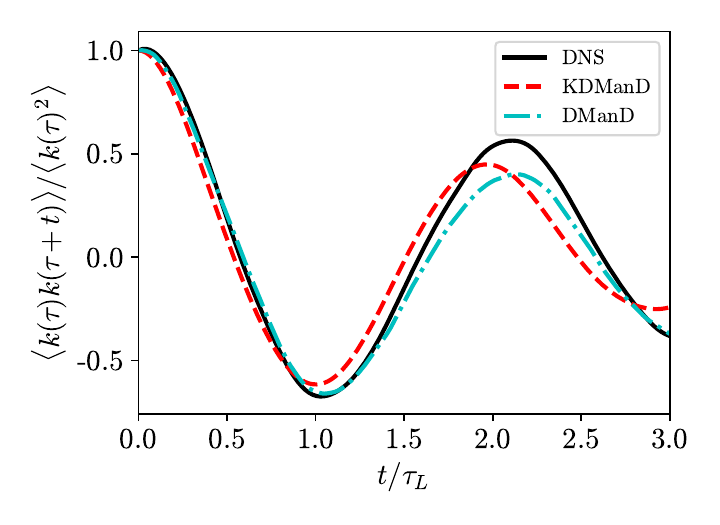} \\
        (a) & (b) \\
    \end{tabular}
    \caption{(a) Normalised short-time-tracking error for 10 random initial conditions using KDManD and DManD up to 150 time units. (b) Temporal autocorrelation of the fluctuations of the kinetic energy for KDManD, DManD and DNS.}
    \label{short_time}
\end{figure}

Now, we apply the state space and function space approaches described in section \ref{framework} (with $\delta t=1$) to learn a linear mapping of the dynamics in the  manifold coordinates, and a nonlinear function, respectively.
\CRCA{We select 200 dictionary elements to find the observable space. Similarly to what was observed for the Koopman MFE model, by using the projection for temporal predictions, described by Eq.~\ref{eq:projection},  we achieve a good approximation of the true dynamics .}

We start the discussion of the predictive capabilities of KDManD and DManD by selecting an initial condition that \MDGrevise{displays the streak breakdown process characteristic of wall turbulence.}
To evaluate how close the models can capture the dynamics of the system, Figure \ref{IC}b represents the kinetic energy of the flow:
\begin{equation}
E_k=\frac{1}{2L_xL_z} \int_{0}^{L_x}\int_{-1}^{1}\int_{0}^{L_z}  \boldsymbol{v}^2 ~ dx~dy~dz.
\end{equation}
We evolve the same initial condition with the DNS solver and in the models. We observe that KDManD can faithfully capture the dynamics, with a good match up to $3\tau_l$ Lyapunov times (here the Lyapunov time corresponds to around $\tau_l \approx 48$ time units, reported by \citep{Inubushi}).
Figure \ref{IC}c  displays field snapshots in the $x-z$ plane ($y=0$) up to $3\tau_L$ for the DNS (top panels) and KDManD (bottom panels). We observe that  streaks (velocity functions below the mean)  at the center of the domain (see $t/\tau_L = 0$) are affected by spanwise wavy disturbances, leading to their breakdown and the formation of rolls (see $t/\tau=0.52-1.04$). These rolls, in turn, induce the re-emergence of streaks (see $t/\tau=2.08$), which break down (see panels $t/\tau=2.39-3.12$) again to restart the sustaining process of turbulence.

While Figure \ref{IC}b,c depicts the trajectory for a single selected initial condition, Figure \ref{short_time}a illustrates t ensemble relative tracking error in the POD coefficients for 500 randomly chosen initial trajectories with KDManD and DManD. Here, the relative error of the models remains below one  until one $\tau_L$, indicating that the models can \CRCA{reasonably} capture the dynamics for approximately one Lyapunov time \CRCA{with approximately $80\%$ of relative error at one $\tau_L$}. We highlight the small error at $t/\tau_L=0$, attributable to the effective coordinate transformation that maps initial conditions from the full space to the manifold coordinates.   Figure \ref{short_time}b illustrates the temporal correlation of the fluctuating part of the kinetic energy, defined as $k(\tau) = E(\tau) - \langle E \rangle$. To generate this figure, we utilized 1000 different initial conditions evolved up to  $3\tau_L$. We observe that KDManD matches well for one $\tau_L$, with small discrepancies persisting up to  $3\tau_L$. 

Next, we focus on the performance of  KDManD and DManD to capture long-time dynamics of the system. Thus, we will evaluate their performance in the reconstruction of the Reynolds stresses, and  the prediction of the energy balance at long times. For the energy balance, we calculate the  power input ($I$) injected due to its moving plates  and the  dissipation due to viscosity ($D$), defined as
\begin{equation}
I=\frac{1}{2L_xL_z}\int_{0}^{L_x}\int_{0}^{L_z}  \left | \frac{\partial v_x}{\partial y}\right |_{y=-1} + \left | \frac{\partial v_x}{\partial y}\right |_{y=1} ~ dx~dz,
\end{equation}
\begin{equation}
D=\frac{1}{2L_xL_z} \int_{0}^{L_x}\int_{-1}^{1}\int_{0}^{L_z} | \boldsymbol{\nabla} \times \boldsymbol{v}|^2~  dx~dy~dz.
\end{equation}
On an energy balance, the energy input rate, the dissipation rate, and the kinetic energy  $E  = (I -D)/Re$,  must average to zero over long times. 

\begin{figure}
    \centering
\begin{tabular}{cc}
\includegraphics[width=0.65\linewidth]{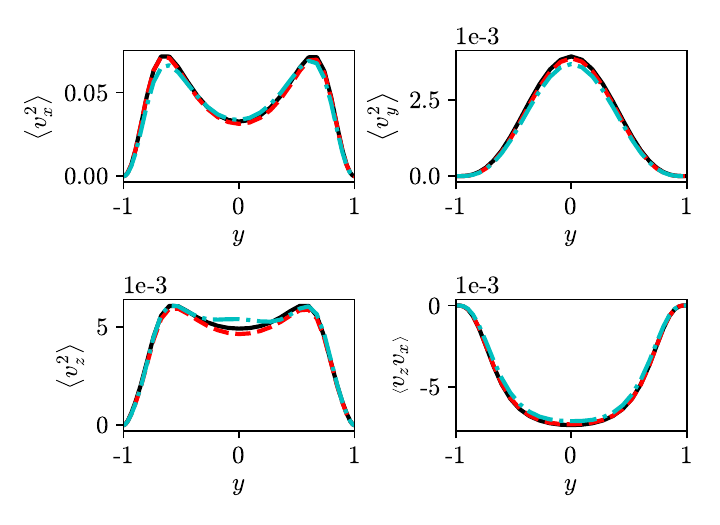} &
\includegraphics[width=0.34\linewidth]{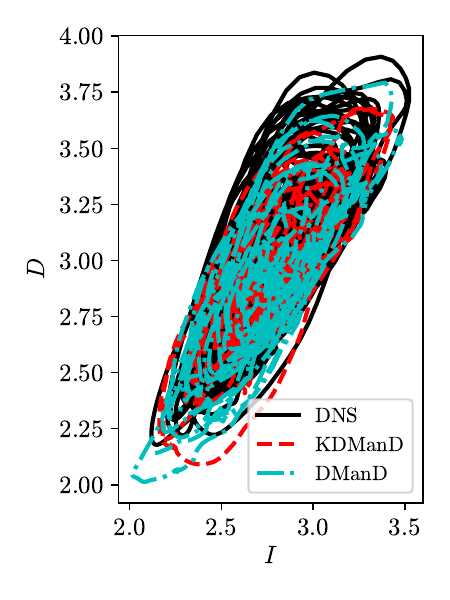} \\
(a) & (b)\\
\end{tabular}
\begin{tabular}{c}
\includegraphics[width=\linewidth]{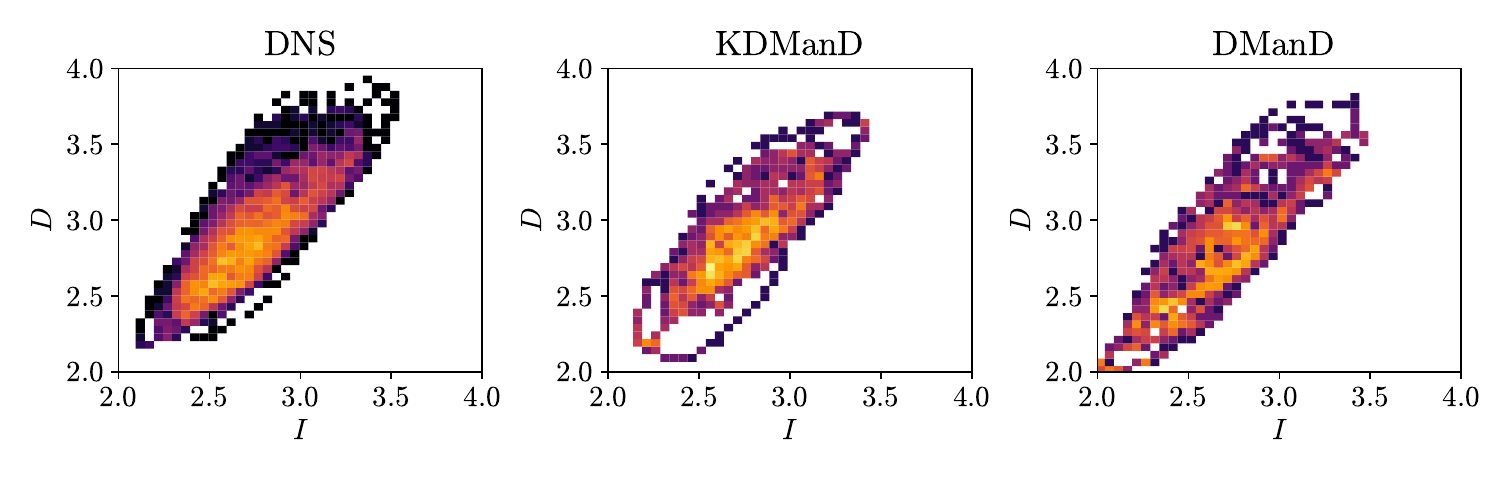} \\
(c)
\end{tabular}
\caption{Long-time statistics of the models and the DNS. The same IC has been evolved for $40\tau_l$ for the models: (a) Components of the Reynolds stress. (b)  Energy dissipation vs power input.  (c) PDF plots of the dissipation vs power input.  }
\label{Long_time}
\end{figure}

Figure \ref{Long_time} summarises the long-time statistics of the DNS and the models for an initial condition (IC) that has been evolved for approximately $40\tau_L$. Figure \ref{Long_time}a displays  four components of the Reynolds stresses, KDManD demonstrates remarkable accuracy in predicting the DNS results, showing a very small deviation from the true solutions, some small discrepancies are observed in DManD model for the $\left \langle  v'^2_z \right \rangle$ component. To ensure that the system effectively captures the energy balance, Figure \ref{Long_time}b displays the $D-I$ projection of the states predicted by KDManD and DManD, with black shading indicating a long trajectory from the DNS solver. It is observed that KDManD and DManD models effectively capture the core region of these projections, with some excursions occurring at low dissipation rates that are not present in the DNS results. 
Finally, Figure \ref{Long_time}c shows the PDF from the DNS (generated from a single trajectory evolved for $100\tau_L$), the PDF of KDManD and DManD models. Overall, we can conclude that KDManD can effectively predict accurately the long-time statistics of this complex system in manifold coordinates. 

Finally, we note that at very long times (beyond 
$40\tau_L$), most KDManD models encounter stability problems. This is a current limitation of the framework presented in section \ref{framework_function}.
We note that this issue might be resolved with a stabilization term, as employed in the DManD framework, but addressing this is beyond the scope of the present research.
\CRCA{We did not observe this stability issue in the MFE model. We believe this is due to the low dimensionality of the system's state space.
}

\section{Conclusions}\label{sec:conclusions}
Simulations of shear flows play a crucial role in the study of turbulence as they provide a simplified setting to explore fundamental mechanisms of turbulence, such as the self-sustaining process of wall turbulence \citep{Hamilton1995}. Most of the works applied for the modelling of shear flows are based on the state space approach. In the present work, we investigate shear flows using both state space and function space approaches. For the latter, we lift the state space to a higher dimensional observable space where \CRCA{ the temporal evolution of the observables can be captured through the Koopman operator.}
To find this set of observables (we use neural networks as `universal approximators') and the corresponding Koopman operator, we employ a variant of EDMD-DL which leverages automatic differentiation to perform gradient descent through the Moore-Penrose pseudoinverse \citep{edmd_dl_ad}. 

First, we applied the Koopman methodology to the MFE model \citep{moehlis2004low}. Since the model is inherently low-dimensional, and only described by nine coefficients, we applied the  the Koopman framework directly in the state space. Our data-driven Koopman model accurately predicts both the short-time and long-time statistics of the system. We compared our results with \citep{Fox2023} (which used the CANDyMan framework presented in \citet{Floryan}), 
and the echo state results presented by \citet{Racca2022}. Our Koopman framework yielded better short-time predictions (accurate for almost 3 Lyapunov times) and long-time predictions, including the reconstruction of Reynolds stresses, the mean velocity profile, and the joint PDF of coefficients  $a_1$ and $a_3$.

Next, we moved to the high dimensional Minimal Flow Unit  of plane Couette flow  at $Re=400$. 
\CRCA{To ensure all the relevant spatial and temporal scales are fully resolved,}
the initial dimensionality of the full state space is ($\mathcal{O}(10^5)$. Lifting this space to a higher observable space for Koopman predictions renders this approach computationally intractable.
But, leveraging the dissipative nature of the Navier-Stokes equations, where viscosity attenuates small-scale perturbations, results in the long time dynamics living in an invariant manifold of lower dimension.  To  identify a coordinate transformation to a manifold representation  of ($\mathcal{O}(10^1)$),  
we introduced the IRMAE-WD-B framework, which provides a consistent estimate of the manifold dimension ranging between  $d_h=32$ and $d_h=36$; we selected the latter dimension for the function space and state space modeling. 
Note that this dimension is larger than the one predicted by \citet{alec_couette}, who found $d_h=18$ by varying the latent dimension and evaluating the mean squared error of the reconstruction using another type of advanced autoencoder \citep{alec_pre}. Then, we lifted the manifold state space to a high-dimensional observable space where the dynamics can be described by the linear Koopman operator. Our Koopman framework successfully captured the \sout{self-sustaining} streak breakdown process, tracked short-term dynamics up to \CRCA{nearly one} Lyapunov time,
and accurately reconstructed the long-term statistics evaluated in the Reynolds stresses and  the power input-dissipation energy balance. Additionally, we included results from state space modeling using neural ODEs to learn the evolution equation in manifold coordinates. We observed comparable results between the function space and state space frameworks.

A promising direction for future research is leveraging the linear representation afforded by the Koopman operator. This linearization simplifies control, enabling the use of established techniques such as the Linear Quadratic Regulator (LQR) for optimal control in nonlinear systems. We are actively developing model-based control strategies within manifold dynamics and extending them to the full state space (similar to the work of \citet{Linot_control}).
Integrating LQR with the Koopman framework could significantly enhance control precision for nonlinear systems by transforming their dynamics into a higher-dimensional linear framework. This approach balances performance and energy usage effectively and could advance applications from fluid dynamics to robotic motion planning.

\section*{Declaration of Interests} The authors report no conflict of interest.

\section*{Acknowledgments}
\noindent This work was supported by ONR N00014-18-1-2865 (Vannevar Bush Faculty Fellowship). We gratefully acknowledge Alec J. Linot for helpful discussions.

\section*{Data availability statement}
Code and sample data that support the findings of this study for finding manifold coordinates with IRMAE-WD-B are openly available at \url{https://github.com/mdgrahamwisc/IRMAE_WD_B}


\bibliographystyle{jfm}
\bibliography{KoopmanCouette.bib}

\end{document}